\documentclass[fleqn,usenatbib]{mnras}

\usepackage{newtxtext,newtxmath}
\usepackage[T1]{fontenc}
\usepackage{dsfont}

\DeclareRobustCommand{\VAN}[3]{#2}
\let\VANthebibliography\thebibliography
\def\thebibliography{\DeclareRobustCommand{\VAN}[3]{##3}\VANthebibliography}

\providecommand{\tabularnewline}{\\}

\newcommand{\vrr}{\vrule width 0pt depth 0pt height 0pt}

\usepackage{bm}

\usepackage{graphicx}	% Including figure files
\usepackage{amsmath}	% Advanced maths commands

\title[Thermal instability with polybaric pressure]{Thermal instability and multiphase dynamics in the ISM with polybaric pressure effects}

\author[H. Sarkar and M.P. Bora]{
Hitendra Sarkar,$^{1}$\thanks{E-mail: hsarkar@gauhati.ac.in}
and Madhurjya P. Bora$^{1}$
\\
$^{1}$Physics Department, Gauhati University, Guwahati 781014, India.
}
\date{Accepted XXX. Received YYY; in original form ZZZ}

\pubyear{2024}

\begin{document}
\label{firstpage}
\pagerange{\pageref{firstpage}--\pageref{lastpage}}
\maketitle
\begin{abstract}
%%%%%%%%%%

%%%%%%%%%		
In this work, we have carried out a two-dimensional (2D) simulation of thermal instability (TI) in interstellar matter (ISM), considering it to be a weakly ionized inviscid plasma with radiation loss. We carry out the simulation using our multi-fluid flux-corrected transport (\emph{m}FCT) code which incorporates a background magnetic field and anisotropic pressure. The anisotropic pressure is modeled with a polybaric pressure model. The findings of our analysis are consistent with the contemporary status of knowledge about  the multiphase nature of the ISM with volume and mass fractions of the various components of the ISM that is warm, cold, and unstable neutral matter (UNM) in the ranges reported by various numerical and observational analysis. Though the strength of the background magnetic field \emph{only} marginally affects the overall evolution, the ratio of the parallel and perpendicular pressures can considerably alter the mass and volume fractions of the three phases, which can affect the overall evolution of the TI in the long run.
\end{abstract}


%%%%%%%%%%%%%%%%%%%%%%%%%%%%%%%%%%%%%%%%%%%%%%%%%%

%%%%%%%%%%%%%%%%% BODY OF PAPER %%%%%%%%%%%%%%%%%%

\section{Introduction} \label{sec:1}

Thermal instability (TI) is a physical process responsible for non-gravitational condensations in various plasma environments such as solar corona, solar wind, magnetosheath, interstellar medium (ISM), intracluster medium (ICM) etc. When thermal equilibrium becomes unstable, it can change the structure and dynamics of the medium at a small scale resulting in condensations of high density and low temperature such as clouds and filaments, known as the precursors to star formation. First introduced by  \citet{Parker_1953} and then by Field in his seminal work \citeyearpar{Field_1965}, TI is known to explain the existence of the multiphase structure of the ISM and mass flow between different intermediate structures, which has very significant effects on the rate of star formation \citep{Chieze_1987}.

The multiphase model of the ISM is a well-studied physical model, which was thought to be consisting of two stable phases -- warm and cold, formed out of neutral atomic hydrogen \citep{Field_1969}. Studies have later confirmed the existence of these two phases -- diffuse warm neutral medium (WNM) and dense cold neutral medium (CNM), believed to be maintained by two thermodynamical states of neutral hydrogen \citep{Wolfire_1995,Wolfire_2003}. Based on earlier studies, it has been argued that large amount of thermally unstable matter is \emph{not} allowed to exist in theoretical models of global ISM \citep{Kalberla_2018} and even if exists, it has to be a transition phase. However, it has been subsequently found, in numerical simulations \citep{Kritsuk_2017,Kim_2023} as well as observationally \citep{Kalberla_2018,Murray_2018,Heiles_2003a,roy_2013} that a large fraction of the ISM is actually in the unstable neutral media (UNM), comprising about $\sim20-40\%$, both in volume and mass fractions, making the ISM actually a three-phase medium. As such, we do not have sufficient physical understanding of the UNM phase, the presence of which definitely affects the status of both WNM and CNM phases. We should note that besides these three phases, there is also a hot ionized phase, known as the hot ionized medium (HIM), existing at a much higher temperature, believed to be maintained by supernov\ae\ heating and a warm ionized phase, maintained by hot B-type stars, known as warm ionized medium (WIM) \citep{McKee_1977}. We in this work however, do not consider the WIM or HIM phase in our analysis.

Starting from Field's work \citeyearpar{Field_1965}, there have been various one-dimensional (1D) studies of TI in both linear and nonlinear regimes with different background conditions. Hennebelle and others \citeyearpar{Hennebelle_1999,Hennebelle_2000} have observed that transonic converging flows of warm gas can produce long-lived (more than cooling time) condensations of cold gas and in the presence of a magnetic field, in which condensation forms along the field direction due to realignment of fluid flow if the angle between original flow and field is small enough. Another concurrent study of shock propagation into WNM and CNM observed fragmentation of thin dense, thermally collapsed layer into molecular clouds \citep{Koyama_2000}. They proceeded with several further simulations of TI in one, two, and three dimensions \citep{Koyama_2000,Koyama_2002,Inutsuka_2005}. There have been many other studies which investigated propagation of a shock front in a thermally unstable gas driving two-phase turbulence in the ISM \citep{Koyama_2002,Audit_2005}, effect of kinematic viscosity on nonlinear turbulent fluid flow \citep{Inutsuka_2005,Koyama_2006}, and nonlinear evolution of TI for converging flows of WNM in presence of a magnetic field \citep{Inoue_2008,Inoue_2009}. \citet{Koyama_2004} included isotropic thermal conduction term and put forward the so-called `Field condition' which limits the simulation cell size to avoid artificial fragmentation of the medium. In a magnetized background, both pressure and thermal conduction become anisotropic, which leads to the formation of long filamentary condensations along the field direction \citep{Choi_2012}. Recently,  \citet{Jennings_2021} have investigated the TI for multiphase gas in the ISM with a Gaussian random field (GRF) density perturbation, in the presence of realistic conduction. Apart from the ISM, there are several other studies on TI in ICM \citep{Parrish_2009,Sharma_2010,McCourt_2011} and solar corona regions \citep{Soler_2011,Antolin_2022,Claes_2019}. \citet{Falle_2020} have reinvestigated Field's linear theory of TI using modern wave interaction and stability analysis, providing a clearer representation of the underlying physical mechanisms. That being so, the study of TI with different initial conditions and new physical processes remain always an interesting area for understanding the balance of cooling and heating in the formation of multiphase structures in various astrophysical environments.

We, in this work perform an inviscid two-dimensional (2D) simulation of the TI in the ISM, with a background magnetic field with pressure anisotropy. We carry out our simulations through a multi-fluid, 2D flux-corrected transport code (\emph{m}FCT), which takes care of any possible numerical diffusion. We use a polybaric pressure model in our simulation, where we handle the pressure anisotropy with an anisotropy parameter $a_{p}$ \citep{Stasiewicz_2004,Stasiewicz_2005,Stasiewicz_2007}. With the polybaric pressure equation, we can model the ISM pressure as polytropic in nature, which is supported by observations \citep{Spaans_2000} in contrast to the widely used Chew-Goldberger-Low (CGL) double adiabatic pressure model \citep{Chew_1956}, which requires very slow plasma flow so that the parallel and perpendicular pressures become completely disconnected, a condition very difficult to be justified \citep{kulsrud}. The pressure anisotropy parameter is assumed to be constant which is also supported by the fact that variable pressure anisotropy would require the magnetic field to be strongly dependent on density, which however is not supported by observational data for the ISM density range $\sim0.1-100\,{\rm cm}^{-3}$ \citep{Troland_1986}. The polybaric model has been successfully applied to study large amplitude magnetosonic oscillations in Earth's magnetosheath region. As far as the kinematic viscosity is concerned, we can find many contemporary simulations with inviscid fluid \citep{Padoan_2016,Piontek_2005,Sharma_2010,Audit_2005,Soler_2011,Inoue_2006,Hennebelle_2007,Kobayashi_2022,Kobayashi_2020,Inoue_2012}.  
%%%%%%%%%%%%%
%{
We also note that supernov\ae\ activity often provides the essential triggering mechanism which can drive the thermal instability. 
We  incorporate the  supernova feedback  (or supernova forcing) activity  \citep{Kim_2017} through a random forcing term in our equations. 
%%%
However, we \emph{must} emphasize that we are only interested in large-scale  structure of the ISM and assume that supernov\ae\ shock fronts have decayed into wide-scale random perturbation with solenoidal and compressional components \citep{Padoan_2016}. Our analysis is to primarily assess the effect of polybaric pressure on thermal evolution of the ISM and to see whether we can arrive at the similar outcome regarding distributions of WNM, UNM, and CNM with time-dependent heating/cooling. 
%%%%
We further note that while  supernova feedback  can drive shock fronts through the ISM; from simulation \citep{Kim_2014} and observational studies \citep{Heiles_2003a} on $21\,\rm cm$ neutral hydrogen emission, it is observed that the so-called spin temperature and kinetic temperature remain almost equal in the CNM and UNM regions indicating near sonic or subsonic turbulent velocity, ruling out existence of any shock front, which is consistent with our assumption  \citep{Kobayashi_2022,Kritsuk_2017,Hennebelle_2007,Saury_2014,Kim_2017}. 
%}
%%%%%%%%%%%%%

The organization of this paper is as follows. In Section \ref{sec:2}, we discuss our plasma model along with description of the polybaric pressure equations. We also discuss in this section, why kinematic viscosity may not be too much relevant in the parameter regime that we have studied. In Section \ref{sec:3}, we describe our numerical model and briefly also explain the numerical strategy that we have adopted. Section \ref{sec:4} is devoted to the analysis of the linear phase of the TI, carried out with a scaled-down version of our code in 1D, the findings of which also serve as benchmarking results for our code. The primary results of our analysis i.e.\ the results of fully nonlinear 2D polybaric simulation of TI are presented in Section \ref{sec:5}, where we also present our important observations and findings in various subsections. Finally, in Section \ref{sec:7}, we conclude.

\section{Ideal MHD equations with radiative heat-loss} \label{sec:2}

The conservative form of the ideal MHD equations, which can be written in general as 
\begin{equation} 
    \frac{\partial{\cal F}}{\partial t}+\nabla\cdot\mathbb{F}=0, 
\end{equation} 
where 
\begin{equation} 
    {\cal F}=(\rho,\rho\bm{v},{\cal E},\bm{B})' \label{eq:columnvec} 
\end{equation} 
is a column vector whose components are mass density $\rho$, momentum density $\rho\bm{v}$, total energy density ${\cal E}$, and magnetic field $\bm{B}$. The quantity $\mathbb{F}$ represents the corresponding flux. The individual equations making up the above model are the continuity, momentum, energy equations, and Faraday's law combined with Ohm's law. 
%%%%%%%
The MHD fluid equations in their compressible form are given by 
\begin{eqnarray} 
    \frac{\partial\rho}{\partial t}+\nabla\cdot(\rho\bm{v}) & = & 0,\\
    \frac{\partial}{\partial t}\left(\rho\bm{v}\right)+\nabla\cdot\left(\rho\bm{v}\bm{v}\right) & = & \bm{J}\times\bm{B}-\nabla\cdot\mathbb{P}
    +{\bm f}_{\rm SN},\\
    \frac{\partial{\cal E}}{\partial t}+\nabla\cdot\mathbb{F}_{\textrm{en}} & = & {\bm v}\cdot {\bm f}_{\rm SN}-\rho\mathfrak{L},\label{eq:en}\\ 
    \frac{\partial\bm{B}}{\partial t}+\nabla\cdot\mathbb{F}_{B} & = & 0, 
\end{eqnarray} 
where $\mathbb{P}$ is the anisotropic pressure tensor, and 
%}%%%
\begin{equation} 
    \bm{J}=\frac{1}{\mu_{0}}\nabla\times\bm{B}. 
\end{equation} 
The second term on the right hand side of Eq.(\ref{eq:en}) is the radiative heat-loss term with $\mathfrak{L}$ as the total heat-loss function (cooling and heating) \citep{Field_1965}. The total energy per unit volume ${\cal E}$ is given by 
\begin{equation} 
    {\cal E}=\frac{1}{2}\rho v^{2}+\frac{p}{\gamma-1}+\frac{B^{2}}{2\mu_{0}}, 
\end{equation} 
where the terms on the right represent respectively the kinetic energy, internal (thermal) energy, and magnetic energy. The corresponding energy flux $\mathbb{F}_{\textrm{en}}$ can be written as 
\begin{equation} 
    \mathbb{F}_{\textrm{en}}=\left(\frac{1}{2}\rho v^{2}+\frac{\gamma}{\gamma-1}p+\frac{B^{2}}{\mu_{0}}\right)\bm{v}-\frac{1}{\mu_{0}}\bm{B}(\bm{B}\cdot\bm{v})+\bm{Q}, 
\end{equation} 
with $\bm{Q}$ denoting thermal conductivity. In writing the energy equation, we have expressed the thermal pressure as the average scalar pressure 
\begin{equation} 
    p=\frac{2p_{\perp}+p_{\parallel}}{3},\label{eq:avp-1} 
\end{equation} 
where $p_{\perp,\parallel}$ are the parts of anisotropic pressure tensor 
\begin{equation} 
    \mathbb{P}=p_{\perp}\delta_{ii}+(p_{\parallel}-p_{\perp})\hat{b}_{i}\hat{b}_{j}.\label{eq:aniso-2}
\end{equation} 
The temperature $T$ is related to pressure through the ideal gas law 
\begin{equation} 
    T=\mu\frac{p}{R\rho}, 
\end{equation} 
where $\mu$ is the mean molecular weight. In the above expression, $\hat{b}_{i}$ are the components of the magnetic field vector $(\bm{B}/B)$. For the combined Faraday's and Ohm's law, we have 
\begin{equation} 
    \mathbb{F}_{B}=\bm{v}\bm{B}-\bm{B}\bm{v}. 
\end{equation}

The supernova feedback term ${\bm f}_{\rm SN}$ is modelled through a tunable random forcing term ${\bm f}_{\rm SN}=\delta(t-t_0)\left(\rho{\bm a}-\left<\rho{\bm a}\right>\right)$ \citep{Kritsuk_2017} to have a velocity field with an adjustable ratio of solenoidal and compressive components, which can be triggered at time $t=t_0$.

\subsection{Generalized continuity equation form} \label{sec:2.1}

{To obtain the numerical results presented in this paper, we have employed the Flux-Corrected-Transport (FCT) scheme (Section \ref{sec:3.1}), which is based upon conservation of fluxes while transporting any fluid quantity across structured grids by suppressing numerical diffusion that may arise in the system from discretization of the differential operators. Implementation of the FCT formalism requires the fluid equations to be put in the form of generalized continuity equations.} The flux-conservative form of the above MHD equations can be expressed in a form of generalized continuity equation with a source term, which will actually be used in our computation. This form can be written as
\begin{equation}
    \frac{\partial{\cal F}}{\partial t}=-\nabla\cdot({\cal F}\bm{v})+S,\label{eq:gce}
\end{equation}
where we split up the total flux term into the flux of the quantity
${\cal F}$ and a corresponding source $S$. The column vector ${\cal F}$ is as given in Eq.(\ref{eq:columnvec}) and the source terms are given by
\begin{eqnarray}
    S & = & \left(0,S_{\textrm{mom}},S_{\textrm{en}},S_{B}\right)',\\
    S_{\textrm{mom}} & = & -\nabla\cdot\left(\mathbb{P}+\frac{B^{2}}{2\mu_{0}}\bm{I}-\frac{1}{\mu_{0}}\bm{B}\bm{B}\right),\\
    S_{\textrm{en}} & = & -\nabla\cdot\left[\left(p+\frac{B^{2}}{2\mu_{0}}\right)\bm{v}-\frac{1}{\mu_{0}}\bm{B}(\bm{B}\cdot\bm{v})+\bm{Q}\right]-\rho\mathfrak{L},\\
    S_{B} & = & -\nabla\cdot(\bm{v}\bm{B}-2\bm{B}\bm{v}).\label{eq:sb}
\end{eqnarray}
Note that the supernova forcing terms are not included in the above expressions for brevity and and can always be added afterwards.

\subsection{Radiative heat-loss with thermal conduction} \label{sec:2.2}

While the radiative heat-loss function $\mathfrak{L}\equiv\mathfrak{L}(\rho,T)$ \citep{Field_1965} is a function of the local density and temperature, the thermal conductivity term, which can be written, in presence of a magnetic field, as
\begin{equation}
    \bm{Q}=-K_{\parallel}\nabla_{\parallel}T-K_{\perp}\nabla_{\perp}T,
\end{equation}
where the directions $\parallel,\perp$ refer to that of the equilibrium magnetic field $\bm{B}_{0}$. For interstellar matter (ISM), we follow a model for $K$ and $\mathfrak{L}$, which are prescribed as \citep{Falle_2020}
\begin{eqnarray}
    K & \simeq & K_{\parallel}=K_{c}T^{1/2},\\
    \mathfrak{L} & = & \frac{\rho}{m_{H}^{2}}\Lambda(T)-\frac{1}{m_{H}}\Gamma,
\end{eqnarray}
where $K_{c}$ is a positive constant and $(\Lambda,\Gamma)$ are respectively radiative cooling and heating functions with $m_{H}$ as the mass of neutral hydrogen atom. These functions can further be expressed as
\begin{eqnarray}
    \Lambda & = & \Gamma\left[a\,e^{f_{1}(T)}+bT^{1/2}e^{f_{2}(T)}\right]\,\textrm{cm$^3$\,erg/s},\label{eq:heat-loss}\\
    \Gamma & = & 2\times10^{-26}\,\textrm{erg}/\textrm{s},
\end{eqnarray}
where $a=10^{7}$ cm$^3$ and $b=1.4\times10^{-2}$ cm$^3$/K$^{1/2}$ are positive constants and  $f_{1,2}(T)$ are certain functions of $T$,
\begin{eqnarray}
    f_{1}(T) & = & -\frac{1.184\times10^{5}}{T+1000},\label{eq:cool1}\\
    f_{2}(T) & = & -\frac{92}{T},\label{eq:cool2}
\end{eqnarray}
where $T$ is measured in Kelvins. We should \emph{note} that the equilibrium cooling function is assumed to be always zero
\begin{equation}
    \mathfrak{L}(\rho_{0},T_{0})\sim0.
\end{equation}

\subsection{Equation of state} \label{sec:2.3}

The last piece of the model is to specify the equation of state. The isotropic equation of state can be written as
\begin{equation}
    \frac{d}{dt}\left(\frac{p}{\rho^{\gamma}}\right)=0.\label{eq:state}
\end{equation}
The anisotropic pressure, represented by Eq.(\ref{eq:aniso-2}), needs however two equations of state. At the simplest, the equations of state can be expressed through CGL double adiabatic laws \citep{Chew_1956},
\begin{eqnarray}
    \frac{d}{dt}\left(\frac{p_{\perp}}{\rho B}\right) & = & 0,\label{eq:cgl1}\\
    \frac{d}{dt}\left(\frac{p_{\parallel}B^{2}}{\rho^{3}}\right) & = & 0.\label{eq:cgl2}
\end{eqnarray}
The double adiabatic equations, however, are quite restrictive and can be used theoretically only when the system has very slow variation along the field lines such that the communication having different behaviour at two different points, even along the lines of force, is very little \citep{kulsrud}. Various theoretical and numerical studies also indicate that to a large extent the equation of state for ISM should be polytropic in nature \citep{Spaans_2000}, which contradicts the CGL equations of state as in CGL, both parallel and perpendicular pressures have different polytropic indices. To circumvent this, the polybaric equations of state are prescribed as \citep{Stasiewicz_2004,Stasiewicz_2005,Stasiewicz_2007}
\begin{equation}
    p_{\perp}\propto N^{\gamma}B^{\kappa},\label{eq:polybaric1}
\end{equation}
along with a pressure anisotropy parameter
\begin{equation}
    a_{p}=p_{\parallel}/p_{\perp}-1,\label{eq:aniso}
\end{equation}
where $\gamma,\kappa$ are certain exponents. This model does not impose any restrictions on adiabacity. In what follows, we shall use the $\gamma=5/3$.

\subsection{Inviscid plasma flow} \label{sec:2.4}

While viscosity is a fundamental characteristic of any fluid including plasma, in many situations the fluid viscosity can be safely neglected which simplifies the underlying computational methods. Many such examples can be found which include studies of ISM, ICM, and intergalactic medium \citep{Padoan_2016,Piontek_2005,Sharma_2010,Audit_2005,Inoue_2006,Hennebelle_2007,Kobayashi_2022,Kobayashi_2020,Inoue_2012}. In any case, the effect of fluid viscosity is negligible in the larger scale, while it begins to become important at smaller scales. For studies involving ISM, such as the present work, the usual parameter of interest for fluid viscosity is the Prandtl number
\begin{equation}
    {\rm Pr}=\frac{\gamma}{\gamma-1}\frac{k_{B}}{m_{H}}\frac{\mu}{K},
\end{equation}
which is the ratio of the kinetic diffusivity to thermal diffusivity, where $k_{B}$ is the Boltzmann constant and $\mu$ is the coefficient of dynamic fluid viscosity. This number is usually fixed at $2/3$, considering the ISM to be mostly monoatomic gas, which fixes the value for $\mu$ depending on the thermal conductivity $K$. In a turbulent flow, the largest scale at which the effect of fluid viscosity becomes important, is usually the Kolmogorov's scale \citeyearpar{kolmogorov_1941} (ideally for incompressible flow)
\begin{equation}
    l_{{\rm kol}}\sim\left(\frac{\nu^{3}}{\epsilon}\right)^{1/4},\label{eq:kolsc}
\end{equation}
where $\nu$ is the kinematic viscosity of fluid given by $\nu=\mu/\rho$ and $\epsilon$ is the average rate of dissipation of kinetic energy per unit mass. As we shall see in Section \ref{sec:5.2.3}, the Kolmogorov's scale in our case can be estimated to be $\sim0.8\,{\rm pc}$, which comes out to be $\gtrsim l_{{\rm kol}}$ for the parameters we have considered in this work. So, we can safely conclude that the effect of viscosity on the results that are obtained through this simulation work is only minimal.

\section{Numerical strategy} \label{sec:3}

\subsection{The method of flux-corrected transport (FCT)} \label{sec:3.1}

In this work, we shall use the method of flux-corrected transport (FCT) to solve our initial value problem of the system of equations given by Eq.(\ref{eq:gce}). The FCT method was first introduced by \citet{Boris_1973}, which solves the Navier-Stokes equations in two stages -- a transport stage and a flux-corrected anti-diffusion stage. We shall however use Zalesak's scheme \citeyearpar{Zalesak_1979} which uses a different limiter to limit the anti-diffusive fluxes. The multidimensional time advancement scheme is implemented through a time-step splitting method known as \emph{split-step} method, which is a second-order accurate method for small enough time steps.

\subsection{Normalization} \label{sec:3.2}

We now normalize all the physical quantities by their respective equilibrium values (denoted with subscript `$0$'). Specifically, we adopt the following normalization
\begin{equation}
    \rho/\rho_{0}\to\rho,\quad v/c_{s}\to v,\quad t/t_{0}\to t,\quad l/l_{0}\to l,
\end{equation}
where $t_{0}$ and $l_{0}$ are the acoustic time and length scales with $c_{s}=l_{0}/t_{0}$. We normalize the pressure as
\begin{equation}
    p/p_{0}\to p,\quad T/T_{0}\to T,
\end{equation}
where
\begin{equation}
    p_{0}=\frac{R}{\mu}\rho_{0}T_{0}=c_{s}^{2}\rho_{0},\quad c_{s}=\sqrt{\frac{RT_{0}}{\mu}}.
\end{equation}
The magnetic pressure $p_{B}=B^{2}/\mu_{0}$ can similarly be normalised by its equilibrium value $B_{0}^{2}/\mu_{0}$ as the magnetic field is normalised by its equilibrium value $B_{0}$
\begin{equation}
    \bm{B}/B_{0}\to\bm{B}.
\end{equation}
The thermal conductivity term is scaled as
\begin{equation}
    \nabla\cdot\bm{Q}\to-\alpha\nabla\cdot\left(T^{1/2}\nabla T\right),
\end{equation}
where
\begin{equation}
    \alpha=\frac{t_{0}T_{0}^{3/2}K_{c}}{p_{0}l_{0}^{2}}
\end{equation}
is the normalization term and all terms are in normalized form. For interstellar matter (ISM), we use $K_{c}\sim10^{6}\,{\rm erg/cm/K/s}$ and for equilibrium pressure we have $p_{0}/k=4\times10^{4}\,{\rm K/cm^{3}}$. For other equilibrium quantities, we take $l_{0}=1\,{\rm pc}$, $t_{0}=1\,{\rm Myr}$, $n_{0}\sim2\,{\rm cm^{-3}}$, $T_{0}\sim500\,{\rm K}$, and $\mu=1.27$. With these parameters, we have $\alpha\sim0.037$.

For $T\sim T_{0}$, a quick calculation shows that $f_{1}\ll f_{2}$, appearing in the cooling functions Eqs.(\ref{eq:cool1},\ref{eq:cool2}). So, in the normalized form, we can scale the radiative heat-loss term as
\begin{equation}
    \rho\mathfrak{L}\to\xi\rho^{2}T^{1/2}f_{2}(T)-\delta\rho,
\end{equation}
with
\begin{eqnarray}
    \xi & = & b\frac{t_{0}T_{0}^{1/2}\rho_{0}^{2}\Gamma}{m_{H}^{2}p_{0}},\\
    \delta & = & \frac{t_{0}\rho_{0}\Gamma}{m_{H}p_{0}}.
\end{eqnarray}
For the parameters mentioned above, $\xi\simeq0.072$ and $\delta\simeq0.23$, which are the normalization terms for the radiative loss function. Note that when normalized, the function $f_{2}(T)$ becomes
\begin{equation}
    f_{2}(T)\simeq e^{-0.184/T}.
\end{equation}
We \emph{must} now note that as all the quantities are normalized with their respective equilibrium values and at equilibrium, we have $(\rho,T,p,B)=1$. We should however note that at equilibrium
\begin{equation}
    \mathfrak{L}(\rho_{0},T_{0})\sim0,
\end{equation}
which actually fixes some of the parameters, depending on the state of equilibrium that is assumed. In any case, one must have
\begin{equation}
    \delta=\xi f_{2}(T_{0})
\end{equation}
in equilibrium.

In this work, we consider two cases -- one with zero magnetic field and the other with a constant background $\bm{B}$. Without loss of any generality, we assume that the equilibrium magnetic field lies in the $x$-$y$ plane 
\begin{equation}
    \bm{B}_{0}=\hat{\bm{x}}B_{0}\cos\psi+\hat{\bm{y}}B_{0}\sin\psi,
\end{equation}
at an angle $\alpha$ to the $x$-direction. In this case, the only effect of $\bm{B}$ is through the anisotropic pressure. We note that with the present notation, for constant $\bm{B}$, we have
\begin{equation}
    \hat{b}_{i}\hat{b}_{j}=\left(\begin{array}{ccc}
    \cos^{2}\psi & \cos\psi\sin\psi & 0\\
    \sin\psi\cos\psi & \sin^{2}\psi & 0\\
    0 & 0 & 0
    \end{array}\right).
\end{equation}
We also note that as our simulation domain is constrained in the $x$-$y$ plane, we shall restrict all variations in the $(x,y)$ domain only.

\subsection{Numerical cycle} \label{sec:3.3}

The nonlinear simulations are run in 2D as mentioned in the previous section, where we use the supernova forcing term ${\bm f}_{\rm SN}$ to drive the perturbation through a randomised velocity field with a mix of solenoidal and compressive components. Note that the initial state is  unstable to the thermal condensation mode. The use of energy equation is desirable for the purpose of conservation of the total energy. Once we determine the total energy, the $p_{\parallel,\perp}$ and the total pressure $p$ can be calculated using Eqs.(\ref{eq:avp-1},\ref{eq:aniso}). The temperature then can be determined from the scalar (averaged) pressure by using the normalised relation
\begin{equation}
    T=\frac{p}{\rho}.
\end{equation}
We use the FCT method to solve Eqs.(\ref{eq:gce}-\ref{eq:sb}) in a uniform grid of size $(1000\times1000)$ with periodic boundary conditions.

In Section \ref{sec:4}, we  present the results for the linear growth phase of the TI in a 1D case. In the 1D case, we use a sinusoidal perturbation, which helps us compare the linear growth phase of the TI with the analytical results \citep{Field_1965}.

\subsection{The characteristic scales} \label{sec:3.4}

At this point, it is useful to calculate the characteristic time and length scales of the problem. Based on the linear instability analysis already outlined before \citep{Field_1965}, one can define the critical wave number for TI to grow, which provides a characteristic length scale, known as the so-called Field length \citep{Falle_2020,Koyama_2004}, which can be defined as
\begin{equation}
    \lambda_{F}=\sqrt{\frac{KT}{n^{2}\Lambda}},\label{eq:flength}
\end{equation}
where $K=K_{c}T^{1/2}$. The {Field} length defines the minimum scale that is required for the TI to grow or else, any density perturbation smaller than the Field length will be quickly erased by the thermal conductivity. Though Koyama and Inutsuka \citeyearpar{Koyama_2000}  used a simplified version of the heat-loss function \citep{Wolfire_1995} to calculate $\lambda_{F}$, one can also use the heat-loss function as given by Eq.(\ref{eq:heat-loss}). For the parameters used in this work, the Field length comes out to be $\lambda_{F}\sim0.24-0.35\,{\rm pc}$ for $T\sim500-1500\,{\rm K}$. As we can see, our computational grid of cell size 
{\begin{equation}
    \Delta x=2\pi\times10/1000\simeq0.063\,{\rm pc}
\end{equation}
}can very effectively resolve the Field length. We also note that our time step size $\Delta t\lesssim0.01$ is quite within the CFL limit
\begin{equation}
    t_{{\rm CFL}}\sim C_{{\rm CFL}}\Delta x/c_{s},\quad C_{{\rm CFL}}<1.
\end{equation}

Next, we calculate the cooling time scale as \citep{Koyama_2004}
\begin{equation}
    t_{{\rm cool}}\sim\frac{kT}{(\gamma-1)}{\rm min}\left(\frac{1}{\Gamma},\frac{1}{n\Lambda}\right)\simeq0.16\,{\rm Myr}.
\end{equation}
This can also be used to define the cooling length of the system as \citep{Koyama_2004}
\begin{equation}
    \lambda_{{\rm cool}}\sim c_{s}t_{{\rm cool}}\sim0.3\,{\rm pc}.
\end{equation}
We can also define a conduction time scale as \citep{Koyama_2004}
\begin{equation}
    t_{{\rm cond}}\sim\frac{nk}{(\gamma-1)K}\Delta x^{2}\sim0.022\,{\rm Myr}.
\end{equation}
We can now safely conclude that our simulation very comfortably resolves the characteristic time and length scales of the problem.

Finally, we do a consistency check by calculating Jean's length, which puts the upper limit for any structure to be gravitationally stable. Though we have not considered self-gravity in our model, calculating Jean's length justifies our \emph{a priori} assumption that gravitational effect may be safely neglected. For parameters already mentioned before, the Jean's length can be calculated as
\begin{equation}
    \lambda_{J}=\sqrt{\frac{\pi c_{s}^{2}}{G\rho}}\sim220\,{\rm pc}.
\end{equation}
As we shall see in Section \ref{sec:5}, considering the highest density condensation of the order of $\sim10\rho_{0}$ and temperature drop of about $\sim0.1T_{0}$ in the CNM, the local Jean's length in the condensations may drop to about
\begin{equation}
    \lambda_{J}^{{\rm local}}\sim26\,{\rm pc},
\end{equation}
both of which are considerably larger than the average condensation size (see Fig.\ref{fig:The-evolution-of}), which is consistent with our assumption of negligible gravitational effect in the model.

{
At this point, we note that the Field length, as determined from the linear dispersion relation can be written in terms of the characteristic wavenumber of the isobaric perturbation due to radiation condensations divided by the average mean free path of conducting particles \citep{Field_1965}
\begin{equation}
k_F^2=k_K(k_T-k_\rho),
\end{equation}
where $k_F$ is the wavenumber corresponding to the Field length, $k_{T,\rho}$ are the wavenumbers of isothermal and isochoric perturbations, and $k_K$ is the reciprocal of the mean free path of the conducting particles. The corresponding Field length is then given by
\begin{equation}
\lambda_{F}\sim k_{F}^{-1}=\left[\frac{\rho_{0}}{K}\left(\mathfrak{L}_{T}-\frac{\rho_{0}}{T_{0}}\mathfrak{L}_{\rho}\right)\right]^{-1/2}.\label{eq:flengthlin}
\end{equation}

The relation given by Eq.(\ref{eq:flength}) \citep{Koyama_2004} is actually obtained by considering isochoric perturbation
\begin{equation}
\lambda_{F}\sim\left(\frac{K}{\rho_{0}\mathfrak{L}_{T}}\right)^{1/2}
\end{equation}
and approximating $\mathfrak{L}=\left(\rho_{0}/m_{H}^{2}\right)\Lambda(T)$
and writing $\mathfrak{L}_{T}\sim\mathfrak{L}/T$.
As we can see that the Field length obtained from linear dispersion relation Eq.(\ref{eq:flengthlin}) diverges near the marginal stability point where $\mathfrak{L}_{T,\rho}\to0$. 
 As such, it should not be interpreted as a literal physical scale in the nonlinear regime, but rather as a guideline for the smallest linearly unstable mode and as a numerical resolution criterion \citep{Koyama_2004}. In our nonlinear simulations, characteristic condensation scales are instead set by the interplay of turbulence and pressure balance. We can apply the same argument for the cooling as well as conduction lengths.
}
{
\subsection{Grid convergence}

The FCT method with  Zalesak's flux limiter \citep{Zalesak_1979} enforces monotonicity by restricting anti-diffusive fluxes. But, it can still retain some numerical diffusion in the process which may cause the solution to evolve into a staircase of square-wave like structure. However, with sufficient grid resolution, this effect can be minimized keeping the underlying physics intact. 
\begin{figure*}%[t]
	%\begin{centering}
	\includegraphics[width=0.49\textwidth]{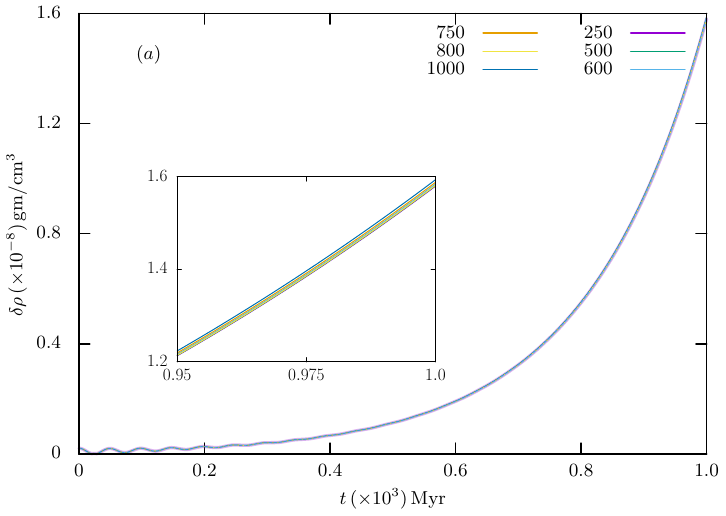}
	%\hfill{}
	\includegraphics[width=0.49\textwidth]{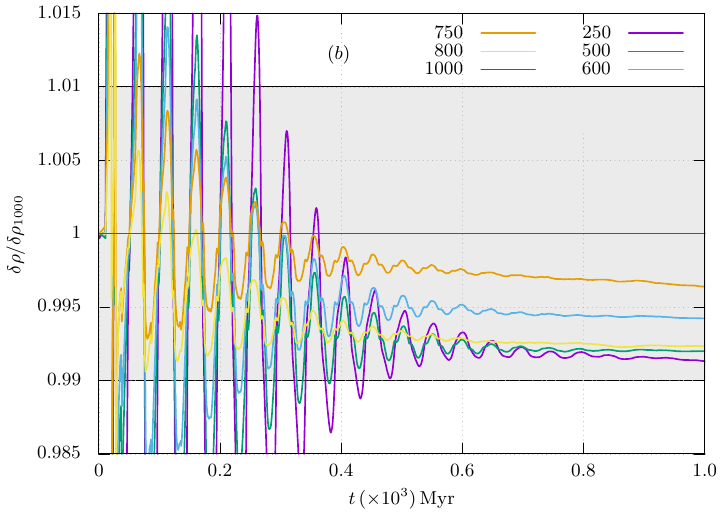}
	\par
	%\end{centering}
	\caption{\label{fig:Grid}(\emph{a}) The temporal density evolution at different grid size is shown. Note that a grid size of $1000$ corresponds to a grid resolution of $0.063\,{\rm pc}$, which is about 10 times smaller than the so-called Field length. The legends in the figure indicate the grid sizes and they all overlap within the accuracy limit. (\emph{b}) Relative density evolution $(\delta\rho/\delta\rho_{1000})$ in time, where $\delta\rho_{1000}$ is the change in density with a grid size of $1000$. The shaded region in the figure shows $\pm1\%$ band around the $\delta\rho_{1000}$ value. We see that relative density for all the grid size (from $250$ onwards till $1000$) fall within this $\pm1\%$ band, indicating a grid-independent evolution.}
\end{figure*}
To check the grid-dependence in our simulation, in Fig.\ref{fig:Grid}, we have shown the temporal evolution of the absolute change in normalised density (left), which corresponds to Fig.\ref{fig:Comparison-of-linear}(\emph{d}) later in Section \ref{sec:4}, and relative density (right) are shown for linear stage of the instability. In Fig.\ref{fig:Grid}(\emph{a}), the temporal density evolution for grid sizes $250, 500, 600, 750, 800$, and $1000$ are shown, which all overlap with each other. The highest grid resolution is for grid size $1000$, which corresponds to a grid resolution of $0.063\,{\rm pc}$ that is about $10$ times smaller than the so-called Field length and the lowest grid size $250$ corresponds to resolution of $0.252\,{\rm pc}$, which is just about the Field length. The legends in the figure show various grid sizes. The right panel shows the relative density deviations ($\delta\rho/\delta\rho_{1000}$) for different grid sizes ($\delta\rho_{1000}$ being the density for grid size $1000$) and the shaded region indicates $\pm1\%$ band around the $\delta\rho_{1000}$. As we can see that all the density values are within the $\pm1\%$, indicating  grid-convergence of the simulation results.
}
\section{The linear growth phase} \label{sec:4}

We know that the linear thermal instability in the absence of magnetic field can essentially grow in two different ways -- the so-called condensation mode and the growing wave mode. As the results are quite well known \citep{Field_1965}, we just mention here the instability conditions without derivation. In the absence of thermal conduction, these instability conditions for condensation and wave modes are respectively the isobaric and isentropic instability conditions, which in the normalized forms can be written as 
\begin{figure*}%[t]
	%\begin{centering}
	\includegraphics[width=0.49\textwidth]{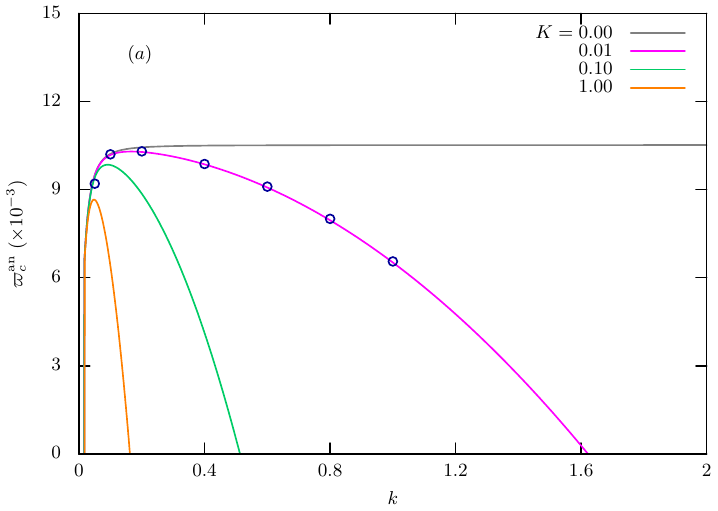}
	%\hfill{}
	\includegraphics[width=0.49\textwidth]{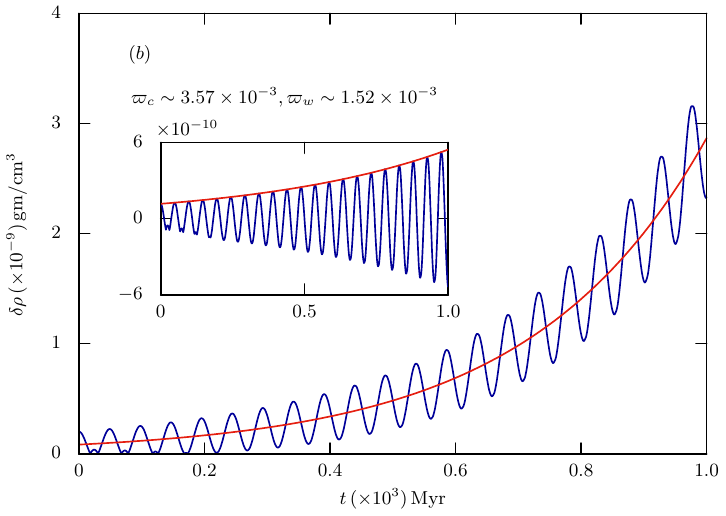}
	%\vfill{}
	\includegraphics[width=0.49\textwidth]{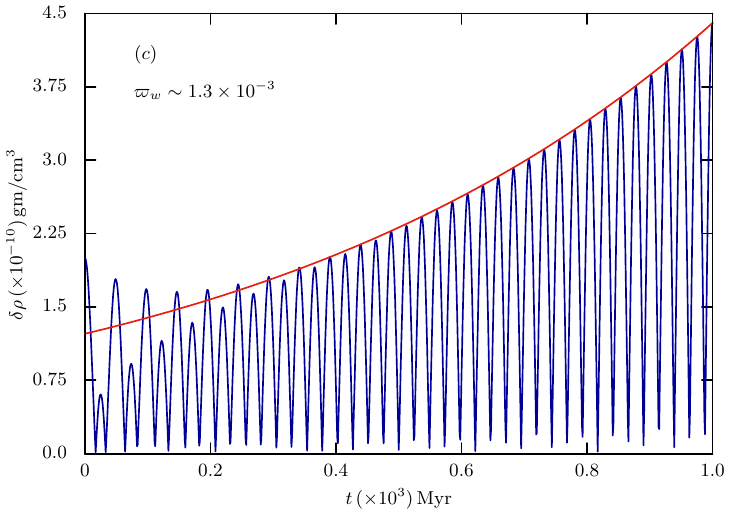}
	%\hfill{}
	\includegraphics[width=0.49\textwidth]{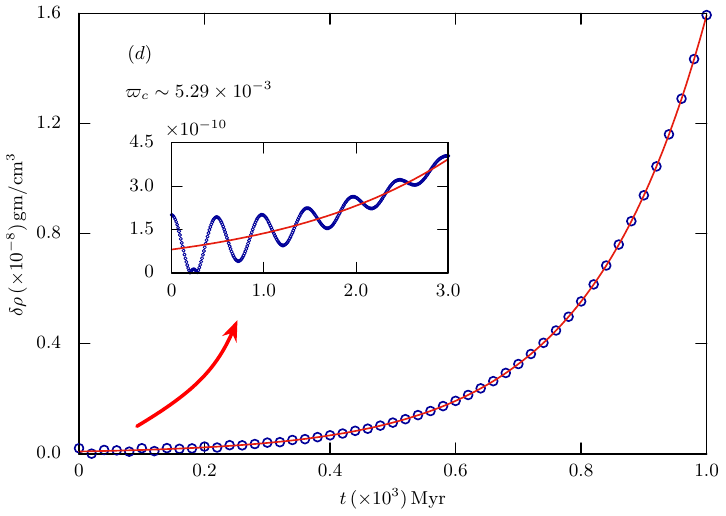}
	\par
	%\end{centering}
    \caption{\label{fig:Comparison-of-linear}{Top left panel (\emph{a}) shows normalized growth rates of the condensation mode for different values of conductivity coefficient (points from simulation are shown by blue circles). Rest of the panels show linear evolution of condensation and wave modes with their growth rates estimated from simulation. (\emph{a}) both modes unstable [panel (\emph{b})], (\emph{b}) only wave mode unstable [panel (\emph{c})], (\emph{c}) only condensation mode unstable [panel (\emph{d})]. The corresponding growth rates $\varpi_{c,w}$ are shown in the figure, as determined from the simulation.} The blue colored graphs (solid and circle) indicate data from simulation and the red colored solid lines indicate an exponential fitting corresponding to the linear growth phase.}
\end{figure*}

\begin{eqnarray}
    \mathfrak{L}_{T}-\mathfrak{L}_{\rho} & < & 0,\\
    \mathfrak{L}_{T}+\frac{1}{\gamma-1}\mathfrak{L}_{\rho} & < & 0,
\end{eqnarray}
where $\mathfrak{L}_{T,\rho}=\partial\mathfrak{L}/\partial (T,\rho)|_{\rho,T}$. With thermal conductivity, these conditions become
\begin{eqnarray}
    k^{2}K+\mathfrak{L}_{T}-\mathfrak{L}_{\rho} & < & 0,\\
    k^{2}K+\mathfrak{L}_{T}+\frac{1}{\gamma-1}\mathfrak{L}_{\rho} & < & 0.
\end{eqnarray}
Theoretically, one can choose different values for $\mathfrak{L}_{T,\rho},K,k$ to make any or all of the above conditions satisfied. For example, for $\gamma=5/3$, we can easily find the following conditions for various instabilities \citep{Field_1965}.

\begin{equation}
    \begin{array}{lcll}
    \textrm{(\emph{a}) both unstable} & : & \mathfrak{L}_{T}<0,|A|<\mathfrak{L}_{\rho}<{\displaystyle \frac{2}{3}}|A|, \\
    && A=k^{2}K+\mathfrak{L}_{T},\\
    \textrm{(\emph{b}) only wave} & : & \mathfrak{L}_{\rho}<0,\\
    &&-A_1<\mathfrak{L}_{T}<-A_1-{\displaystyle \frac{5}{2}}\mathfrak{L}_{\rho}, \\
    && A_1=k^{2}K-\mathfrak{L}_{\rho},\\
    \textrm{(\emph{c}) only condensation} & : & \mathfrak{L}_{\rho}>A,  A>0,\\
     &  & \mathfrak{L}_{\rho}>{\displaystyle \frac{2}{3}}|A|,  A<0.
    \end{array}\label{eq:conditions}
\end{equation}

{To benchmark our numerical scheme, we use the linear dispersion relation  
\begin{eqnarray}
	\omega^3-\gamma \omega k^2+i
	(\gamma-1)\left[\omega^2\left(k^2 K+ \mathfrak{L}_{T}\right)\right.\nonumber\\
	\left.-k^2\left(k^2K+\mathfrak{L}_{T}-\mathfrak{L}_{\rho}\right)\right]&=&0 \label{eq:diseq}
\end{eqnarray}
by considering a perturbation of the form $\sim e ^{-i(\omega t-k x)}$.

In Fig.\ref{fig:Comparison-of-linear}, we compare the theoretical growth rates calculated using the linear dispersion relation with the growth rate measured from 1D TI simulation. In the same figure, the evolution of a single wave for all the above three conditions along with the corresponding numerically estimated growth rates $\varpi_{c,w}$ of the condensation and the wave modes are also shown. In all the panels of Fig.\ref{fig:Comparison-of-linear}, the blue colored curves (solid and circles) denote the results from simulation and the red colored solid line shows an exponential fitting for the linear growth phase. In Fig.\ref{fig:Comparison-of-linear}(\emph{a}), normalized growth rates of the condensation mode calculated from the linear dispersion equation for different values of conductivity coefficient $(K=0,0.01,0.1\,\rm{and}\, 1)$ are shown along with growth rates calculated from simulation for $K=0.01$. Fig.\ref{fig:Comparison-of-linear}(\emph{b}) shows the evolution of a single wave when both wave and condensation modes are unstable. While the primary figure shows the evolution of the condensation mode with a growth rate of $\varpi_{c}\sim3.57\times10^{-3}$ (normalized), the inset  shows the evolution of the wave mode with a growth rate of $\varpi_{w}\sim1.52\times10^{-3}$, determined from the actual growth of amplitude of the wave after removal of the condensation growth. Fig.\ref{fig:Comparison-of-linear}(\emph{c}) shows the evolution of the wave mode (displayed as a difference between the maximum and minimum amplitudes of the wave over time) with a growth rate $\varpi_{w}\sim1.3\times10^{-3}$. Finally, Fig.\ref{fig:Comparison-of-linear}(\emph{d}) shows the evolution of the condensation mode with a growth rate of $\varpi_{c}\sim5.29\times10^{-3}$. The inset shows a zoomed in portion of the initial evolution, where one can clearly see a stable wave mode with only the condensation mode growing. The analytically calculated growth rates $\left(\varpi_{c,w}^{{\rm an}}\right)$ for the three cases from Eq.(\ref{eq:diseq}) are respectively (\emph{a}) $\varpi_{c}^{{\rm an}}\sim3.56\times10^{-3}$, $\varpi_{w}^{{\rm an}}\sim1.52\times10^{-3}$, (\emph{b}) $\varpi_{w}^{{\rm an}}\sim1.31\times10^{-3}$, (\emph{c}) $\varpi_{c}^{{\rm an}}\sim5.27\times10^{-3}$. As we can see, the growth rates determined from the simulation are almost in complete agreement with the analytically calculated ones.}

\section{Nonlinear polybaric evolution} \label{sec:5}

In this section, we are going to present our primary simulation results with polybaric pressure under a constant magnetic field. All the simulations are carried out in a 2D geometry with a simulation box size of $(20\pi\times20\pi)\,{\rm pc}^{2}$. We use the supernova forcing term as a source of perturbation which is fed into the simulation at the initial stage $t_0=0$ through a randomised tunable velocity field with pre-determined ratio of solenoidal and compressive components.

\subsection{Supernova-driven velocity field} \label{sec:5.1}

We note that supernov\ae\ explosions create a radial flow with an expanding shock into the immediate vicinity. However, as time progresses, at larger scale  these expanding shocks begin to create vorticity and generate turbulence which becomes akin to random fluctuations, which is also supported by MHD simulations \citep{korpi_1999}. So, we assume that the supernova forcing can be approximated as random fluctuation so far as the large-scale ISM structure is concerned \citep{korpi_1999,Padoan_2016}.

So, in order to model the supernova-driven energy injection, we use a velocity field with a mix of solenoidal and compressive parts \citep{Padoan_2016}. In particular, we use three scenarios --- Case\,(\emph{i}): a velocity field with a relatively dominant compressive part $(\sim54\%)$, which can be thought to be like a remnant of supernova shock-driven field, Case\,(\emph{ii}): one with a dominant solenoidal field $(\sim72\%)$ and weak compressive part $(\sim28\%)$, and Case\,(\emph{iii}): one with a  solenoidal part $(\gtrsim95\%)$ only with a negligible amount of compressive part.   We shall observe that a compressive velocity field helps drive the TI relatively quicker in reaching a nonlinear saturation  compared to one with a dominant solenoidal part \citep{com-sol}, though statistically all velocity fields result in similar volume and mass fractions of multiphase ISM components, as shown in Subsection \ref{sec:6}.

\begin{figure*}
    \begin{centering}
    \includegraphics[width=0.52\textwidth]{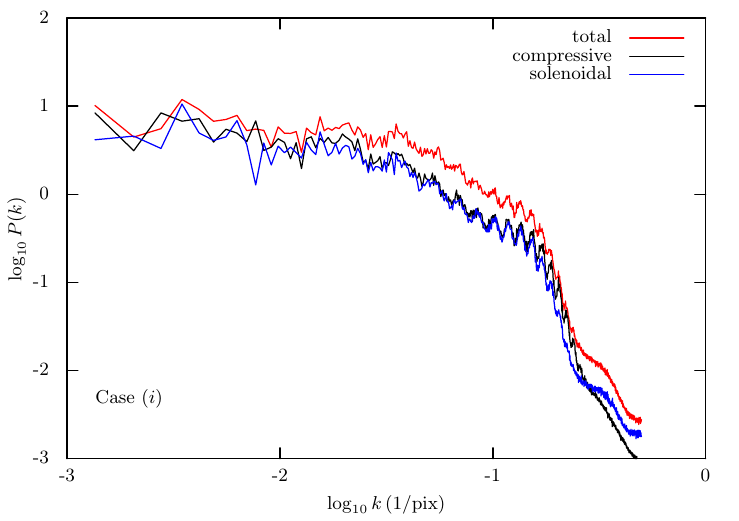}
    \end{centering}
    \caption{\label{fig:sol0}The power spectrum density (PSD) of the initial velocity field used for modelling the supernova forcing term. This velocity field has a relatively dominant compressive part.}
\end{figure*}

In Fig.\ref{fig:sol0},  we have shown the decomposed power spectrum density (PSD) of the net initial velocity for case (\emph{i}). The velocity field has been decomposed through a Helmholtz-Hodge decomposition \citep{Bhatia_2013}
\begin{equation}
    \bm{v}=-\nabla\phi+\nabla\times A,
\end{equation}
where the $-\nabla\phi$ is the compressive part of the velocity field and the other solenoidal. In what follows, we shall analyse our results for this case and in Section \ref{sec:6}, we shall consider the other two cases.

\subsection{No magnetic field} \label{sec:5.2}

\begin{figure*}
    %\begin{centering}
    \includegraphics[width=0.52\textwidth]{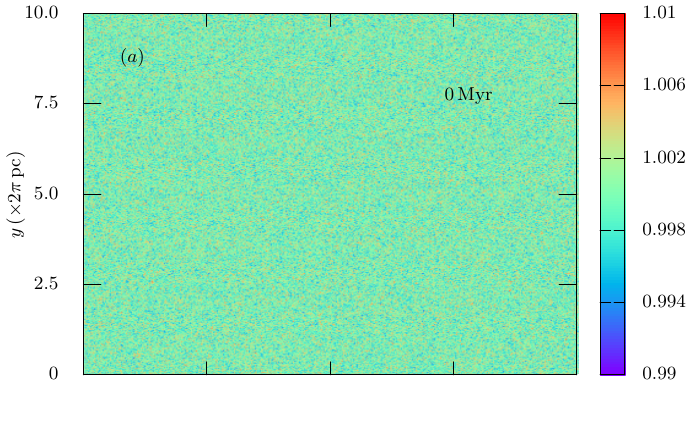}\hskip-12pt
    \includegraphics[width=0.5\textwidth]{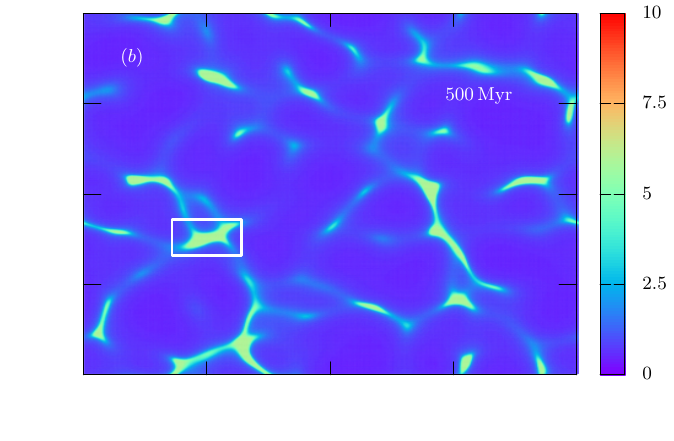}\\
    \vskip-20pt
    \includegraphics[width=0.52\textwidth]{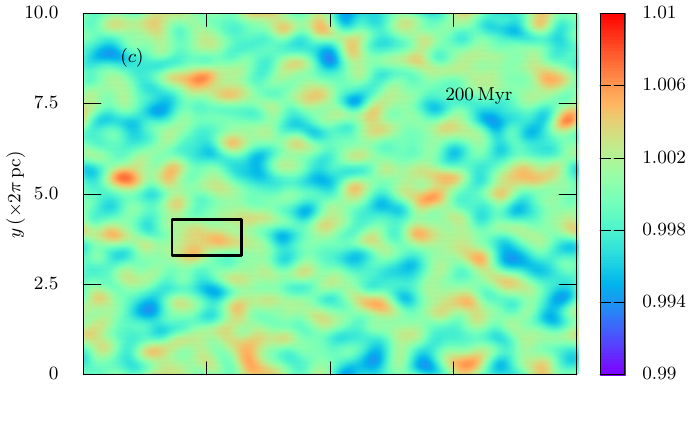}\hskip-6pt
    \includegraphics[width=0.49\textwidth]{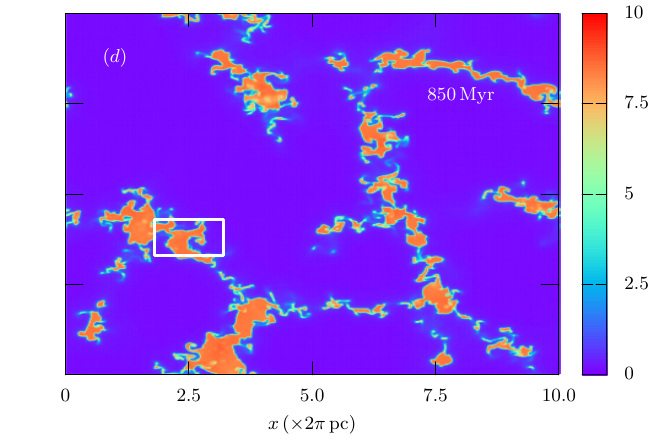}\\
    \vskip-5pt
    \includegraphics[width=0.52\textwidth]{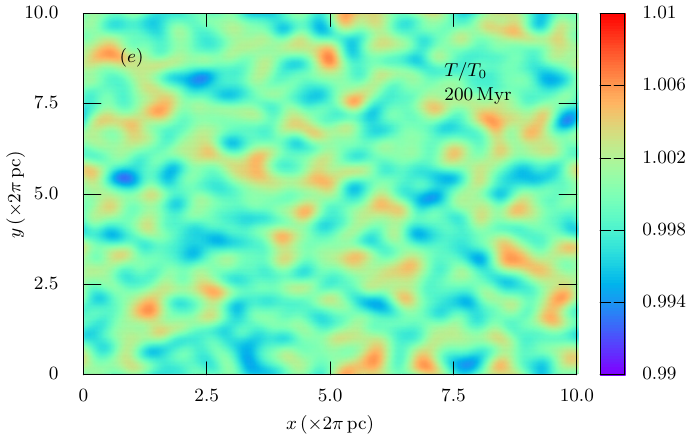}\hskip0pt\raisebox{7pt}{\includegraphics[width=0.44\textwidth]{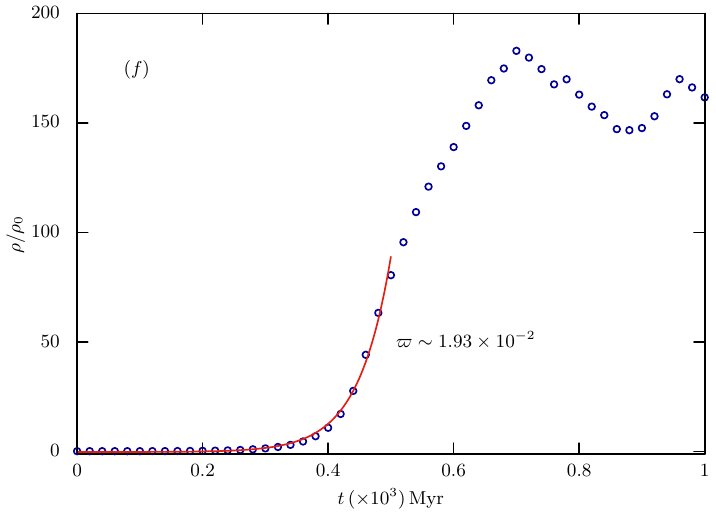}}\hfill{}\mbox{~}
    \par
    %\end{centering}
    \caption{\label{fig:The-evolution-of}The evolution of the density map under TI. The timestamp in different simulation panels are inscribed within each panel. The initial state of the simulation is a random density perturbation, resulting out of supernov\ae\ forcing (see text), which is unstable to the thermal condensation mode. {The small rectangle shown in each of the top four panels indicates a density condensation area, which is used to calculate the growth rate $(\varpi)$, shown in panel (\emph{f}). Panel (\emph{e}) shows temperature perturbations corresponding to density perturbations at $200\,{\rm Myr}$ in panel (\emph{c}).}}
\end{figure*}

In Fig.\ref{fig:The-evolution-of}, we have shown the evolution of the density map with scalar pressure from a random velocity field shown in case (\emph{i}) above. The plots show the thermal condensation in density resulting in formation of cold neutral matter (CNM). The small rectangular box in each panel indicates the area, which is used to calculate the initial (linear) growth rate of the instability. The linear growth of the TI is shown in {Fig.\ref{fig:The-evolution-of}(\emph{f})}. The blue circles indicate the density evolution (within the rectangular box) and the red curve is an exponential fit corresponding to the linear regime. For the ISM parameters already mentioned before and the equilibrium state that we have used, the maximum linear growth rate (with respect to perturbation wave number) that is possible, comes out to be $\varpi_{{\rm max}}^{{\rm theory}}\sim1.03\times10^{-2}\,{\rm Myr^{-1}}$, which we are expected to observe during a simulation with random perturbation. As shown in {Fig.\ref{fig:The-evolution-of}(\emph{f})}, the linear growth rate in this case is found to be $\varpi^{{\rm simul}}\sim1.93\times10^{-2}\,{\rm Myr^{-1}}$, which closely agrees with the theoretical value. The density $\rho$ then increases from an equilibrium value of unity (normalized to the equilibrium value $\rho_{0}$) to about an average value of $\sim180$ during the nonlinear phase. 
As one can see, the nonlinear evolution starts at about $\sim500\,{\rm Myr}$ from a very small value. Naturally, the instability will only linearly grow in the initial phase. We should note that if a higher perturbation amplitude is introduced at the initial stage, the nonlinear evolution will occur sooner. As such, the onset of nonlinear evolution and hence the time required to attain the final condensation stage is quite arbitrary and may depend on the particular simulation concerned. What is however invariant is the linear growth rate of the instability $\varpi$, which is quite consistent with the analytical estimate, as already shown in the previous section and in the beginning of this section. We also note that the nonlinear evolution never attains a completely steady state and fluctuates around a mean value and hence the size and structure of the CNMs will fluctuate with time but the overall density structures have a nonlinear evolution timescale of $\sim10\,{\rm Myr}$. Note that the density shown in this plot is a cumulative density within that rectangle. In {Fig.\ref{fig:The-evolution-of}(\emph{e})}, we have also shown the normalized temperature map $(T/T_{0})$ at a timestamp of $200\,{\rm Myr}$, which clearly shows the anti-correlation between the corresponding density map and formation of CNM and warm neutral matter (WNM), which further confirms the process of thermal condensation with pressure equilibration.

%\newpage
\subsection{Constant magnetic field and pressure anisotropy} \label{sec: 5.2}

We note that experimental observations suggest that the magnetic field strength is almost independent of density in the range of $0.1-100\,{\rm cm^{-3}}$ or in other words the density depleted regions (warm neutral matter or WNM) and the condensations (cold neutral matter or CNM) have roughly the same field strengths \citep{Troland_1986}. We therefore restrict our analysis to a constant background magnetic field $\bm{B}\equiv B\hat{\bm{x}}$ in the $x$-direction. Following \citet{Stasiewicz_2004,Stasiewicz_2005,Stasiewicz_2007}, we also introduce a pressure anisotropy parameter
\begin{equation}
    a_{p}=p_{\parallel}/p_{\perp}-1,
\end{equation}
so that depending on the value of the parameter $a_{p}$, one can have different values for $p_{\parallel,\perp}$.

%\newpage
\subsubsection{Plasma $\beta$} \label{sec:5.2.1}

As the extent of different plasmas of astrophysical interest is quite large, the plasma $\beta$,
\begin{equation}
    \beta=\frac{p}{B^{2}/(2\mu_{0})},
\end{equation}
which is the ratio of the plasma pressure to the magnetic pressure, also can have an extremely wide range of values. While the interstellar matter (ISM) has an extremely wide range of $\beta\sim9\times10^{-8}-7\times10^{8}$ \citep{Haverkorn_2013,Ponnada_2022}, the intracluster medium (ICM) as such have range $\beta\sim0.1-10^{7}$ \citep{Kunz_2011,Berlok_2021}. The magnetosheath region has a $\beta\sim0.1-10^{5}$ \citep{Stasiewicz_2004,Pang_2022,Bandyopadhyay_2022}, while the solar corona and solar wind have a typical range of $\sim10^{-4}-10^{3}$ \citep{Gary_2001,Gomez_2019,Huang_2023}. We have therefore restricted our study to values of $\beta=10^{4}$ and $5$, which we refer to as the `weak' and the `strong' field.

\begin{figure*}%[t]
    %\begin{centering}
    \includegraphics[width=0.53\textwidth]{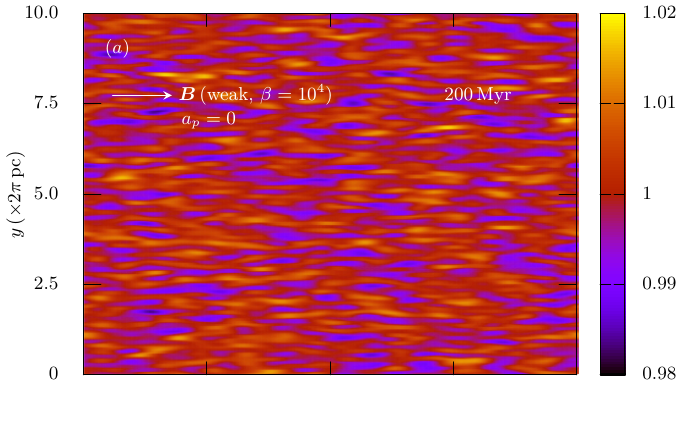}\hskip5pt
    \includegraphics[width=0.455\textwidth]{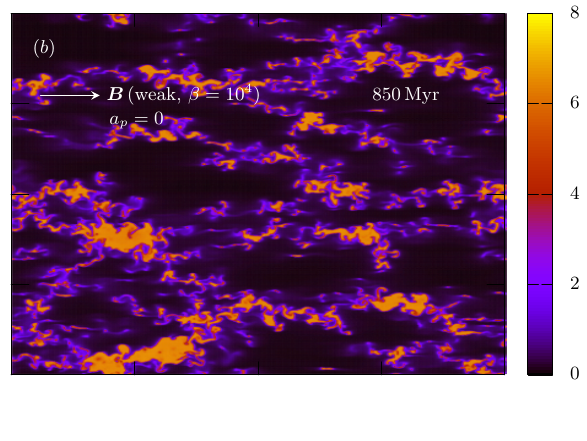}\\
    ~\vskip-36pt
    \includegraphics[width=0.51\textwidth]{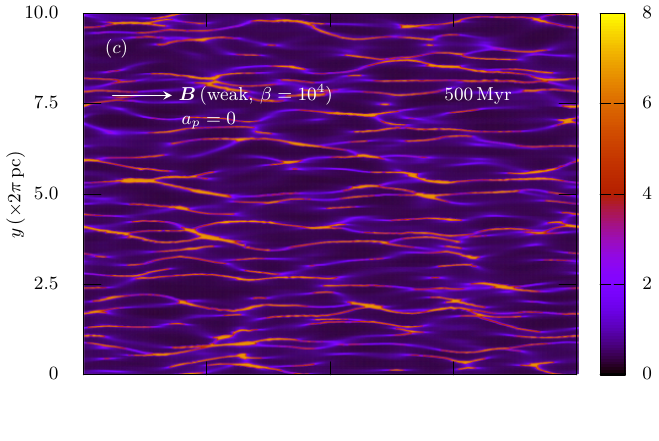}\hskip-12pt
    \includegraphics[width=0.51\textwidth]{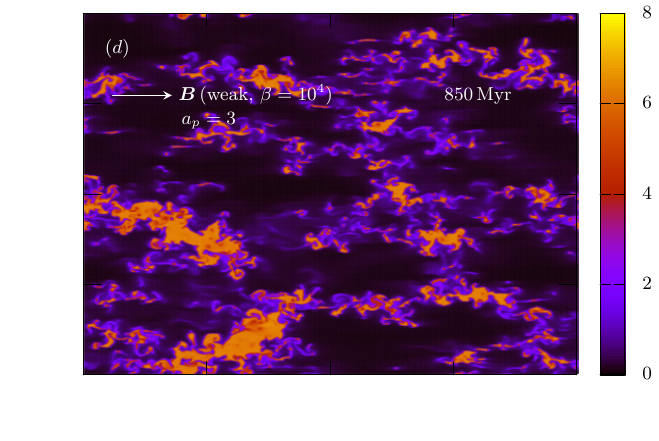}\\
    ~\vskip-36pt
    \includegraphics[width=0.51\textwidth]{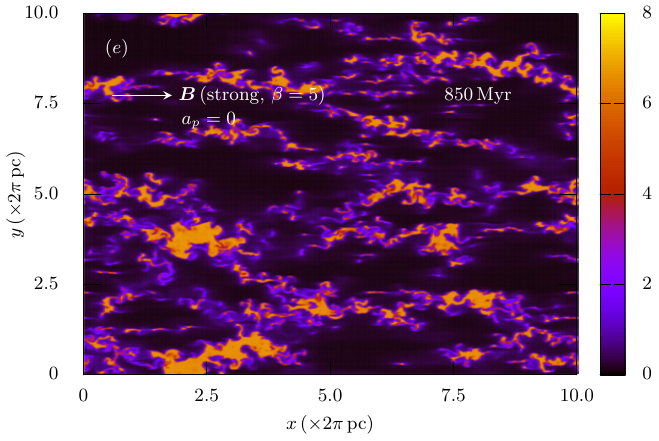}\hskip-12pt
    \includegraphics[width=0.51\textwidth]{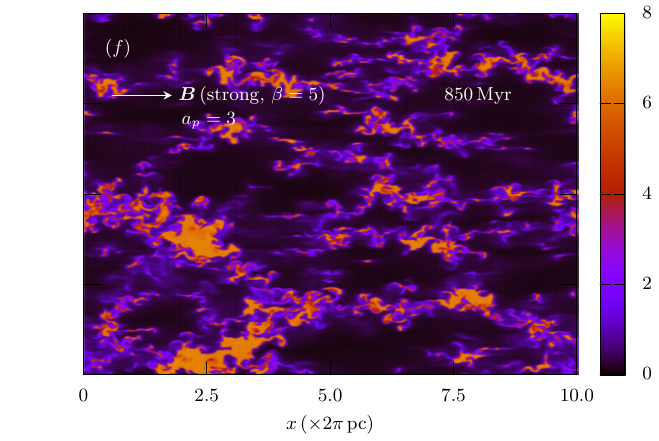}
    \par
    %\end{centering}
    \caption{\label{fig:Density-field-for}Density map for $a_{p}=0$ and $\beta=10^{4}$ (`weak' field) for different phases of the TI [{panels (\emph{a})--(\emph{c})}]. While the {first} three panels show the evolution for $a_{p}=0$, { panel (\emph{d})} shows the same for $a_{p}=3$. The bottom row [{panels (\emph{e}) and (\emph{f})}] shows the evolution at the end of $850\,{\rm Myr}$ for $a_{p}=0$ and $3$ for the case of `strong' field $(\beta=5)$.}
\end{figure*}

\begin{figure*}%[t]
	%\begin{centering}
	\includegraphics[width=0.5\textwidth]{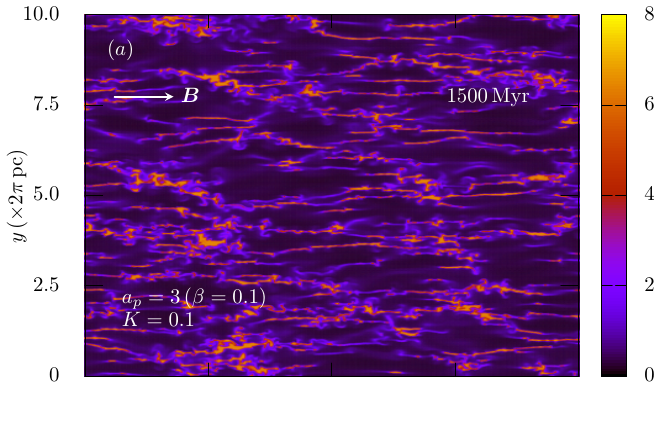}\hskip-6pt
	\includegraphics[width=0.5\textwidth]{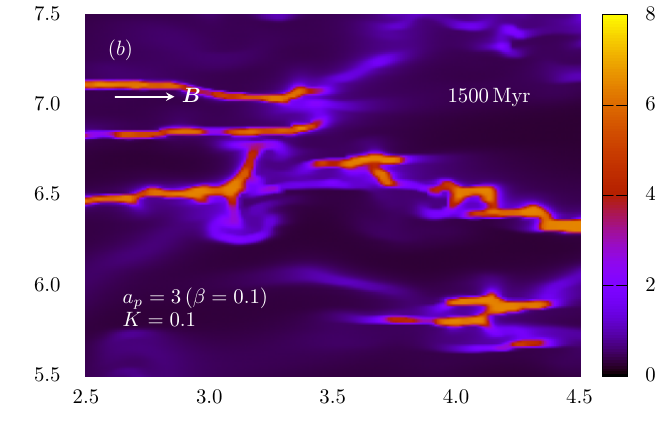}\\
	~\vskip-26pt
	\includegraphics[width=0.5\textwidth]{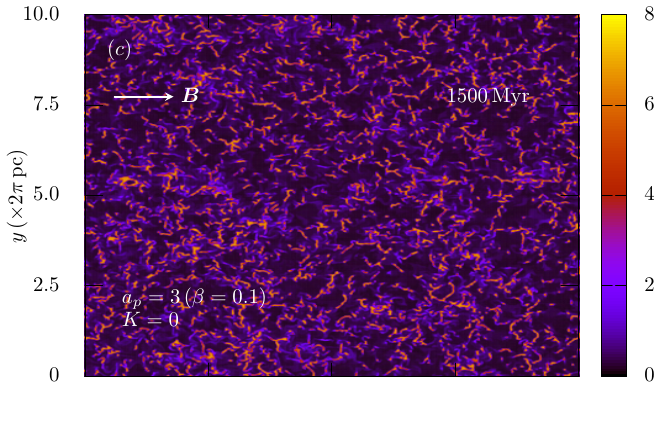}\hskip-6pt
	\includegraphics[width=0.5\textwidth]{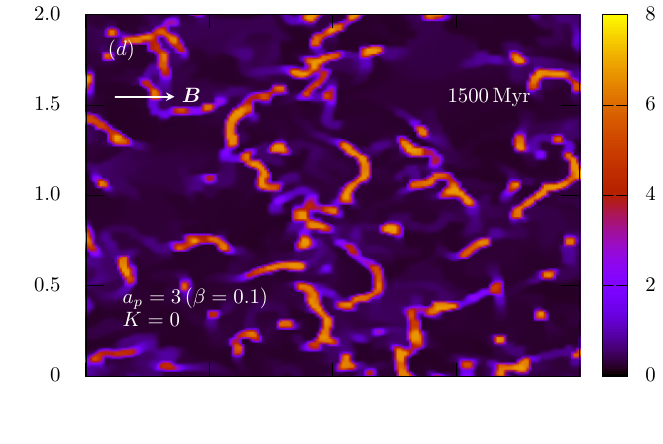}
	~\vskip-26pt
	\includegraphics[width=0.5\textwidth]{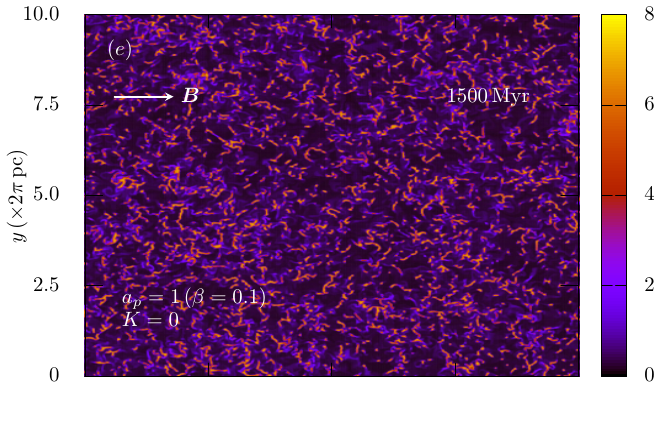}\hskip-6pt
	\includegraphics[width=0.5\textwidth]{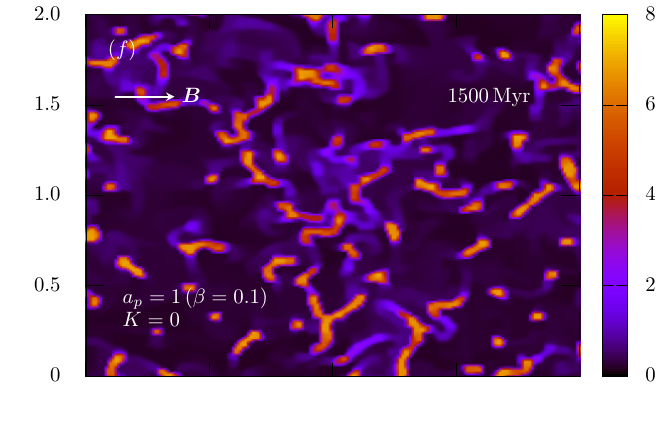}
	~\vskip-26pt
	\includegraphics[width=0.5\textwidth]{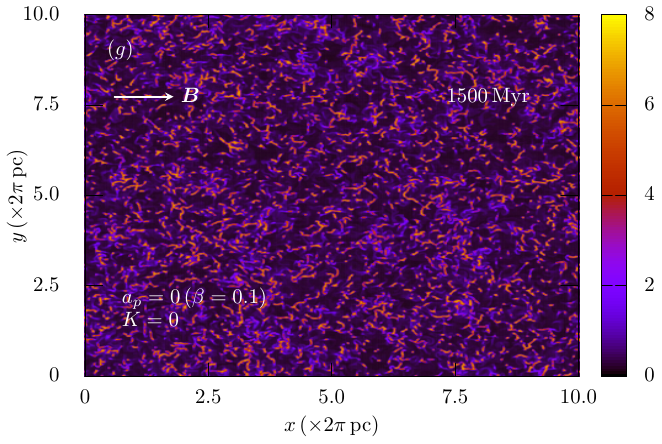}\hskip-6pt
	\includegraphics[width=0.5\textwidth]{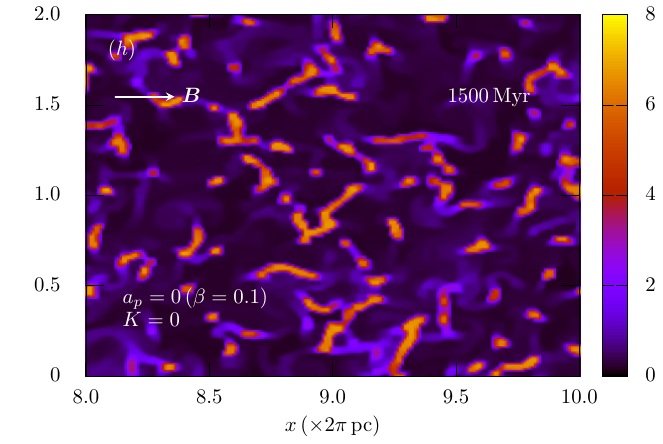}
	\par
	%\end{centering}
	\caption{\label{fig:beta0.1}{Density map for $\beta=0.1$ (`very strong' field) in the presence of thermal conduction with $a_{p}=3$ [panels (\emph{a}) and (\emph{b})] and in the absence of thermal conduction with different values of pressure anisotropy parameter ($a_{p}=0,1,\,\rm{and}\,3$) [panels (\emph{c})--(\emph{h})]. The Right panels show a zoomed in view of the density maps in the corresponding left panels}.}
\end{figure*}

\begin{figure*}%[t]
    \begin{raggedright}
    \includegraphics[width=0.506\textwidth]{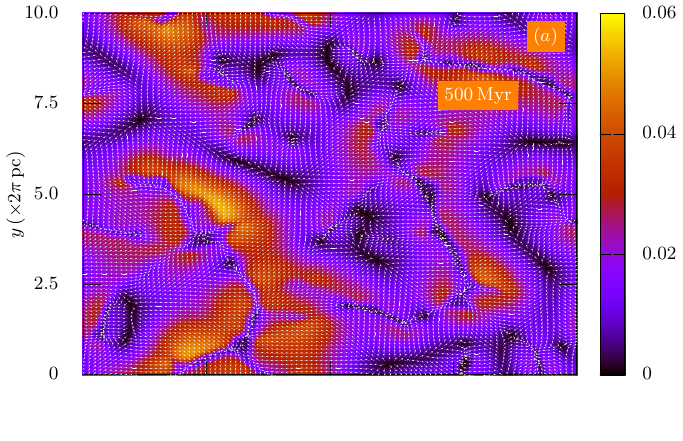}\hskip-6pt
    \includegraphics[width=0.5\textwidth]{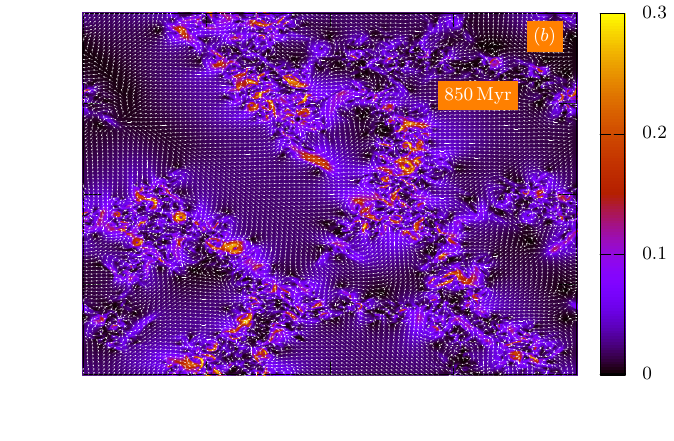}\\
    ~\vskip-36pt\includegraphics[width=0.506\textwidth]{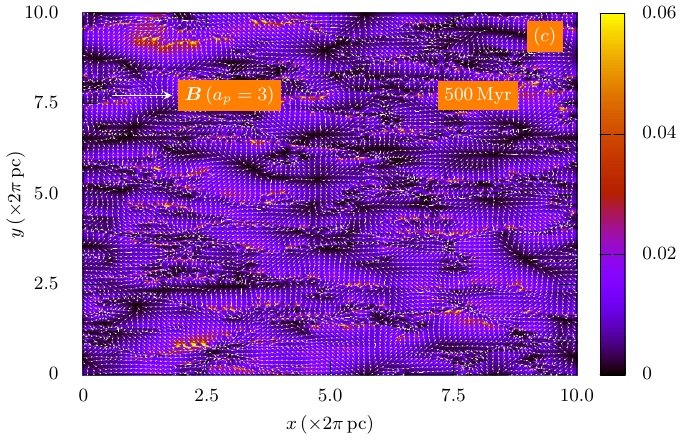}\hskip-6pt
    \includegraphics[width=0.5\textwidth]{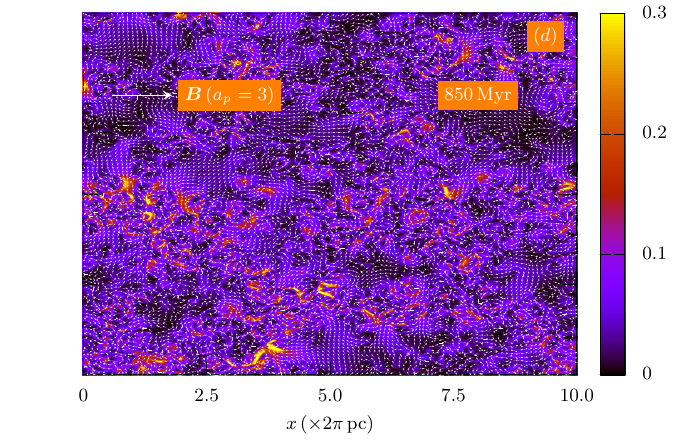}\\
    ~\vskip-12pt
    \includegraphics[width=0.506\textwidth]{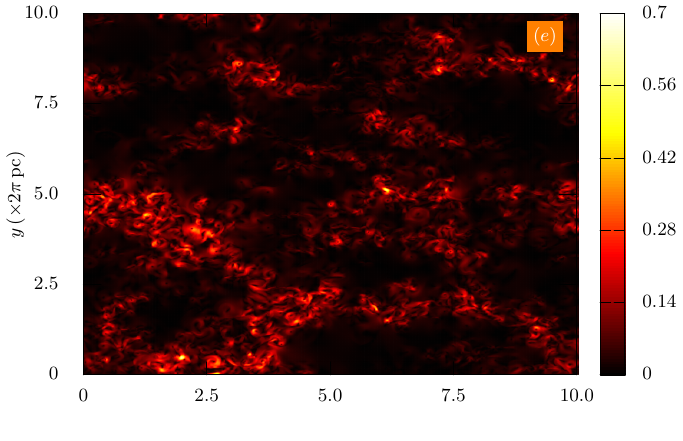}\hskip6pt\raisebox{2pt}{\includegraphics[width=0.44\textwidth]{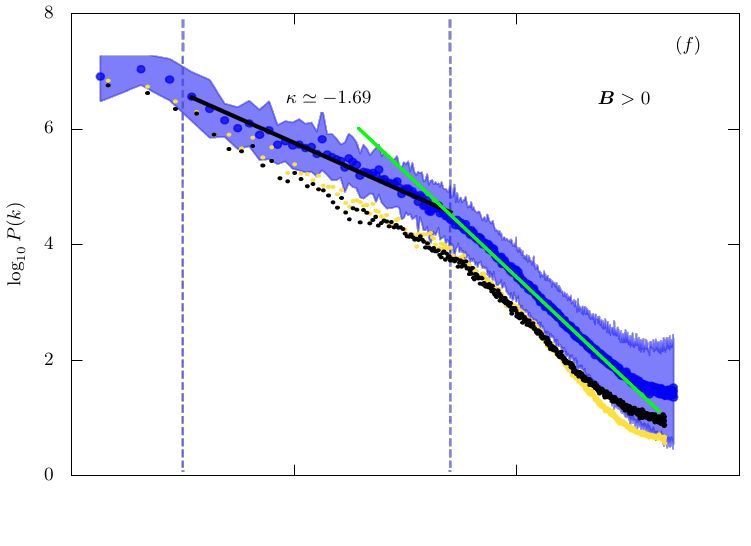}}\\
    ~\vskip-24pt\hskip4pt
    \includegraphics[width=0.45\textwidth]{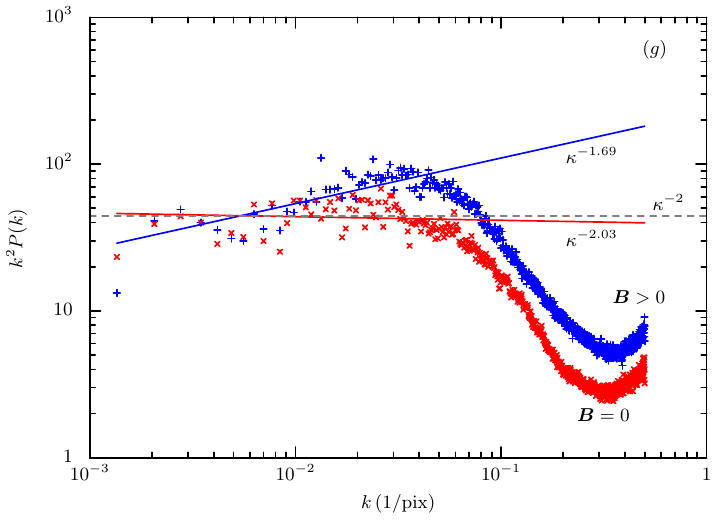}\hskip28pt~\raisebox{2pt}{\includegraphics[width=0.44\textwidth]{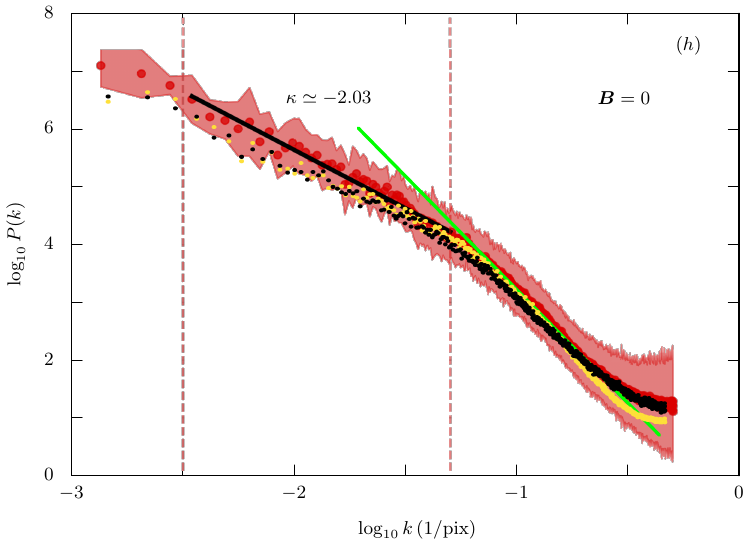}}
    \par
    \end{raggedright}
    \caption{\label{fig:Velocity-field.}The velocity field for $\bm{B}=0$ [{panels (\emph{a}), (\emph{b})}] and `strong' field with $\beta=5$ and $a_{p}=3$ [{panels (\emph{c}), (\emph{d})}] at different time stamps. The local Mach number of the plasma flow for the case $\beta=5$ at $850\,{\rm Myr}$ is shown in {panel (\emph{e})}. The bottom two panels on the right show the power spectrum density (PSD) of the velocity field for $\bm{B}>0\,(\beta=5)$ [{panel (\emph{f})}] and $\bm{B}=0$ [{panel (\emph{h})}]. The respective scalings with the spatial wave number are inscribed in the panels. {Panel (\emph{g})} shows the compensated PSDs of the two cases. The yellow and black colored spectra in both panels indicate the solenoidal and compressive parts that comprise the total spectra.}
\end{figure*}

\subsubsection{Thermal condensation} \label{sec:5.2.2}

In Fig.\ref{fig:Density-field-for}, we have shown the evolution of the density map for the `weak' field $(\beta=10^{4})$ for the parameter $a_{p}=0$ and $3$ (top four panels). The initial state is the same as before (as in the case of scalar pressure). What we can see from the figure is that the initial evolution of density is predominantly along the ambient magnetic field and the final condensations are also developed parallel to the magnetic field. However, as the evolution becomes fully nonlinear, the condensations become diffused compared to the similar map for scalar pressure (see Fig.\ref{fig:The-evolution-of}) though the final condensations {seem to be} aligned with the ambient magnetic field. While all the top four panels in Fig.\ref{fig:Density-field-for} refer to the `weak' field case for $a_{p}=0$ and $3$, the lowermost panel of Fig.\ref{fig:Density-field-for} shows the evolution of density after $850\,{\rm Myr}$ for $a_{p}=0$ and $3$ for the `strong' field case. We can see that for comparatively higher $p_{\parallel}$ (larger $a_{p}$), we have slightly lower condensations.

We now analyse what exactly happens in the case of a strong field. In particular, we choose the case for $\beta=5$ and $a_{p}=3$. We have chosen to plot the velocity field vectors in Fig.\ref{fig:Velocity-field.} which is superimposed on a velocity field map to highlight the plasma flow. Note that in Fig.\ref{fig:Velocity-field.}, the lengths of the velocity vectors are kept uniform for clarity, while the background velocity field indicates the strength of the net plasma flow at that point. The top row of Fig.\ref{fig:Velocity-field.} shows the velocity field vectors and map for scalar pressure when there is no ambient magnetic field. We can see that in this case, the plasma velocity is more or less ordered, driven by the density condensations. During the fully developed nonlinear stage of the evolution at $t\sim850\,{\rm Myr}$, the bulk plasma flow is quite ordered except at the site of the condensations. As we can see that the density condensations are also closely followed by higher plasma flow into the regions of higher mass densities. In contrast to this, when there is a magnetic field present, initially most of the plasma flow is along the magnetic field lines. The low density areas are marked by small-magnitude  plasma flow, as expected. This also causes initial density striations to form along the magnetic field lines. We note that the immediate effect of a magnetic field is to restrict the plasma flow in the transverse direction, which points to the fact that condensation always occurs along the magnetic field lines as the parallel velocity is mostly unrestricted. However, even a small transverse velocity component can make a considerable difference in the final condensation state \citep{Hennebelle_1999}. As a result, we can expect that the thermal instability will be stifled due to the flow restrictions imposed by the magnetic field. However, as the total energy content must be conserved, we certainly expect a field aligned, more diffused condensation state and the resultant velocity field will not be highly ordered leading to a turbulent-like plasma flow, as can be seen from the right figure of the second row of Fig.\ref{fig:Velocity-field.} (see next Subsection).

{We now consider the case for a `very strong' magnetic field with $\beta=0.1$, and show the evolution of density after $1500\,\rm{Myr}$ in the presence of thermal conduction with $a_p=3$ (first row panels) as well as in the absence of thermal conduction for three different values of pressure anisotropy parameter, $a_p=0,1,\,\rm{and}\,3$ (rest of the panels) in Fig.\ref{fig:beta0.1}. For each left panel, the corresponding right panel presents a zoomed-in view of a selected region from the density map.} {In the presence of conduction with a `very strong' field, the nature of initial evolution of density remains the same as in the previous cases with condensations largely elongated in the direction parallel to the magnetic field and  a slower start of the growth of instability. However, as it approaches the nonlinear saturation stage, some of the condensations start getting elongated in the direction perpendicular to the field, which is shown by a zoomed in portion in Fig.\ref{fig:beta0.1}(\emph{b}). In contrast to this, in the absence of thermal conduction, condensations are observed to be very different. These filaments are comparatively shorter, less diffused and get elongated mostly in the perpendicular direction of the field as shown in Fig.\ref{fig:beta0.1}(\emph{c}) and Fig.\ref{fig:beta0.1}(\emph{d}). Similar nature of the condensations in the absence of thermal conduction was earlier reported by \citet{wareing_2016}. The inclusion of anisotropic thermal conduction in the model predominantly channels the formation of condensations along the direction of background magnetic field and suppresses growth along the direction perpendicular to the field \citep{Choi_2012}.

As we understand now, in absence of thermal conduction, there is no mechanism to erase temperature variations. In this case, apparently a strong magnetic field greatly reduces motion of the plasma across field lines, allowing density perturbations along the field lines which can cause sharp density gradient along the field lines resulting filamentary structures which appear to form across field lines \citep{wareing_2016}. In contrast to these, a large parallel thermal conductivity damps temperature perturbations along the field lines. So, small $k_\parallel$ (large parallel wavelength) modes will be less stabilised and will grow fast, which results in field-aligned structures. As we have seen, in both cases our simulation produces the desired results.

At this point, we would like to also emphasise on the effect of pressure anisotropy on the condensations, which is more pronounced for a strong magnetic field. We recall that in our case, the pressure anisotropy can be modelled with a single parameter $a_p$ -- while $a_p=0$ corresponds to $p_\parallel=p_\perp$, a positive $a_p$ corresponds to stronger $p_\parallel$.  Let us consider the panels (\emph{d}, \emph{f}, \emph{h}) of Fig.\ref{fig:beta0.1} (right column) which show the corresponding zoomed-in portion of the density filaments with zero thermal conductivity for $a_p=3, 1$, and $0$ (top to bottom). What we see is that as pressure anisotropy becomes stronger (larger $p_\parallel$), the filamentations become longer, more connected and get elongated across the magnetic field line, while for lower anisotropy, we have scattered and more or less isotropic filamentation. This is a stark deviation from earlier observations \citep{wareing_2016}, for lower $a_p$. Going by our earlier argument, we see that stronger parallel pressure suppresses compressions along the field lines.
}

\subsubsection{Velocity fluctuation and isobaric turbulence} \label{sec:5.2.3}

As our system is dissipation-less, we do not expect to see any shock front and the velocity is supposed to stay subsonic with respect to the local sound velocity at any point. This should not however be confused with supernov\ae\ shocks, which we have assumed to have decayed into random fluctuations, considering the large-scale structure of the ISM that we are interested in.  In {Fig.\ref{fig:Velocity-field.}(\emph{e})}, we show the local Mach number $M=v/\sqrt{T}$, the ratio of the net velocity to that of the local sound speed at a given point within the computational domain. As we can see, the local Mach number always stays well below $1.0$ in the entire computational domain and the plasma flow remains subsonic.
%%%%%%%%%%%%%
%{
We can compare this to the so-called turbulent Mach number $M_t$, determined through observation and defined as
\begin{equation}
M_t=\left[4.2\left(\frac{T_{k,\rm max}}{T_s}-1\right)\right]^{1/2},
\end{equation}
where $T_{k,\rm max}$ is maximum kinetic temperature and $T_s$ is the spin temperature which is basically the  HI excitation temperature. Recent simulation studies \citep{Kim_2014} as well as in observational analysis \citep{Heiles_2003a} show that for  temperature ranges relevant for the CNM and WNM, the ratio $T_{k,\rm max}/T_s\gtrsim1$, which indicates a near sonic or subsonic turbulent velocity, which is consistent with our findings. It should also be noted that a violent initial perturbation can push this Mach number close to unity making a shock-like interface between the CNM and WNM.
%}
%%%%%%%%%%%%%

We note that the basic nature of thermal condensation mode is isobaric and the resultant nonlinear state should also show the nature of isobaric turbulence. We analyse the nature of these turbulence through a spatial power spectrum density (PSD) of the velocity field for both zero and strong magnetic field. In Fig.\ref{fig:Velocity-field.}, we show this spatial PSD for the case $\bm{B}>0,\beta=5$ [{Fig.\ref{fig:Velocity-field.}(\emph{f})}] and for $\bm{B}=0$ [{Fig.\ref{fig:Velocity-field.}(\emph{h})}] with respect to the spatial frequency (expressed as the inverse of the pixel of the velocity field image). What is shown is the 1D power spectrum density of the velocity field calculated using the Turbustat package \citep{Koch_2019}. Note that for the 1D PSD, we have $P(k)\equiv E(k)\propto k^{-\kappa}$, where various values of the index $\kappa$ determine the nature of the turbulence. In the figure, the velocity PSDs are shown in {Figs.\ref{fig:Velocity-field.}(\emph{f}, \emph{h})}, respectively for the case when $\bm{B}>0$ (with $a_{p}=3$) and $\bm{B}=0$. In both these panels, the blue and red colored spectra (with relative error region marked) show the 1D PSD with a {suggested} fitting (thick, solid, black line). The yellow colored spectra in both panels show the solenoidal part of the spectrum and the black colored part shows the compressional part (see the next paragraph for explanation). {Fig.\ref{fig:Velocity-field.}(\emph{g})} shows the compensated spectra for both cases (blue for $\bm{B}>0$ and red for $\bm{B}=0$) for more clarity.

A few points should be noted while interpreting the power spectra shown in Fig.\ref{fig:Velocity-field.}. First, as our system is dissipation-less, the entire range of the power spectra in principle, is inertial as the so-called viscous dissipation range is practically zero. Which is why also, the spectra do not develop any marked transitions indicating the dissipation range. The spectra however show two distinct scalings with a \emph{knee}, one which is marked with black solid line and the other with a green solid line in the figures. We believe that the existence of the latter one is due to the development of numerical viscosity in smaller scales. The so-called inertial range is shown in the figure by two vertical dashed lines. 
These PSD's are with an initial velocity field having the ratio of power of the solenoidal part $(P_{\rm sol})$ to the compressive part $(P_{\rm com})$  as $(P_{\rm sol}:P_{\rm com})=(46:54)$, which is our Case\,(\emph{i}).  The final PSD for $\bm{B}>0$, however shows an almost equal solenoidal and compressive parts with a ratio of $(P_{\rm sol}:P_{\rm com})=(53:47)$.  It is interesting to note that while the scalar pressure (zero $\bm{B}$) velocity field shows a scaling more akin to that of Burger with $E(k)\propto k^{-2}$, with a strong magnetic field $(\beta=5)$, the turbulence scales more close to the classical Kolmogorov scaling with $E(k)\propto k^{-1.69}$. Though the final stage has almost equal amounts of solenoidal and compressive powers in velocity, as expected from a weakly compressible component having low turbulent Mach number (see Fig.\ref{fig:Velocity-field.}), the solenoidal mode is expected to dominate \citep{Elmegreen-2004}, which is why we see an incompressible Kolmogorov-like scaling for the total PSD.

So, with a magnetic field,  the TI-induced isobaric turbulence has grown to a nonlinear saturation with an almost equal amounts of solenoidal and compressive parts in velocity, with slightly more power in the solenoidal part, at least at larger scales. It is also worthwhile to note that recent observations by Voyager I, which indicate incompressible-like turbulence in the very local interstellar medium (VLISM) \citep{Zhao_2020}. We also observe that in the presence of an ambient magnetic field, the final velocity PSD exhibits a Kolmogorov-like scaling, suggesting a transition toward more incompressible turbulence regime. As mentioned, the turbulence remains largely isobaric in nature, which is confirmed by the pressure distribution function, shown in the bottom panel of Fig.\ref{fig:PDFs-for-}. The $\delta$ function-like behaviour of the pressure distribution function indicates that pressure remains isobaric.

As already mentioned in Section \ref{sec:2.4}, our flow is inviscid. In that sense, the entire spectrum range in Fig.\ref{fig:Velocity-field.} can be assumed to be inertial from which we can estimate an equivalent of Kolmogorov's scale to be $l_{{\rm equiv}}\sim0.72\,{\rm pc}$ corresponding to a maximum wavenumber $k\sim0.55\,{\rm pix}^{-1}$. For a Prandtl number ${\rm Pr}=2/3$, the dynamic viscosity is about $\mu\sim0.072$ for the parameters that we have considered in this work. The average rate of dissipation of kinetic energy per unit mass $\epsilon$ can be estimated to be
\begin{equation}
    \epsilon\sim\frac{\left(\overline{u_{{\rm rms}}}\right)^{3}}{l_{{\rm scale}}}\sim5.5\times10^{-7}\,{\rm erg/gm/s},
\end{equation}
where $u_{{\rm rms}}$ is the r.m.s\ velocity of the plasma flow and $l_{{\rm scale}}\sim0.063\,{\rm pc}$ is the smallest scale that is resolved in our simulation. With this, the Kolmogorov's scale {[}Eq.(\ref{eq:kolsc}){]} is estimated to be
\begin{equation}
    l_{{\rm kol}}\sim0.66\,{\rm pc},
\end{equation}
which is $<l_{{\rm equiv}}$, the smallest structure that is generated in our simulation. So, we can see that viscosity plays only a minimal role in determining the overall structure formation due to TI.

\subsubsection{Probability density function (PDF)} \label{sec:5.2.4}

\begin{figure*}
    %\begin{centering}
    %~\hskip4pt
    \includegraphics[width=0.495\textwidth]{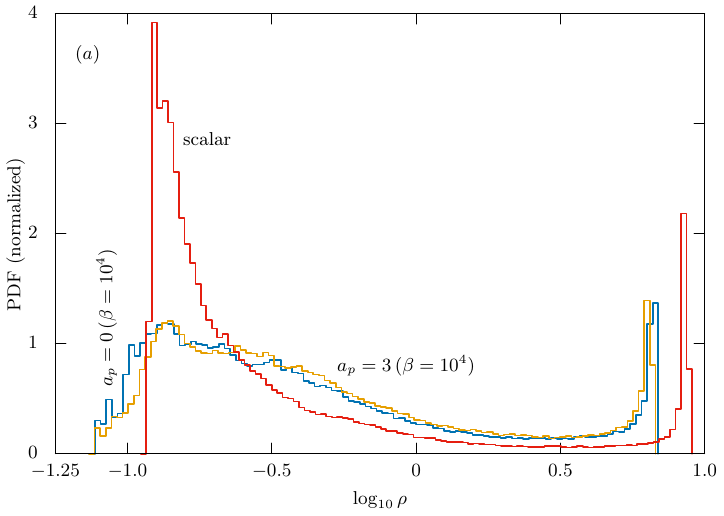}\raisebox{-2pt}
    {\includegraphics[width=0.5\textwidth]{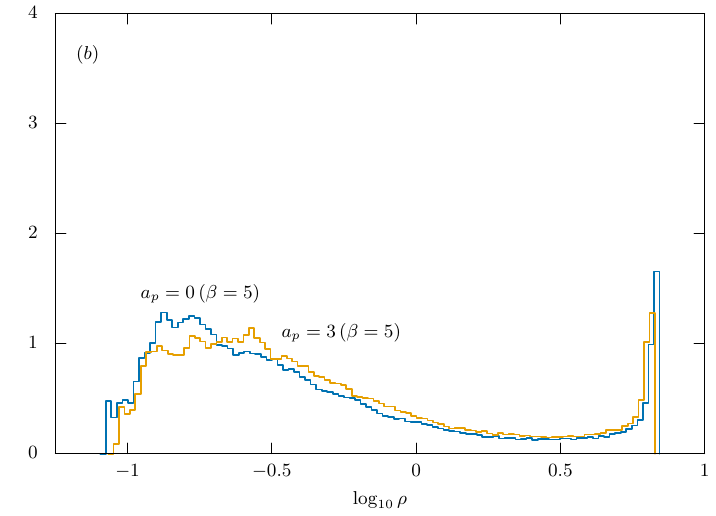}}\\
    \includegraphics[width=0.5\textwidth]{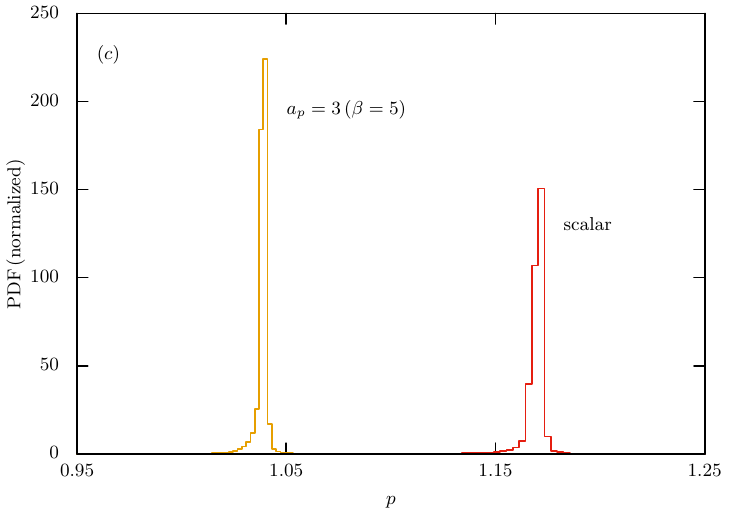}\raisebox{-1pt}
    {\includegraphics[width=0.5\textwidth]{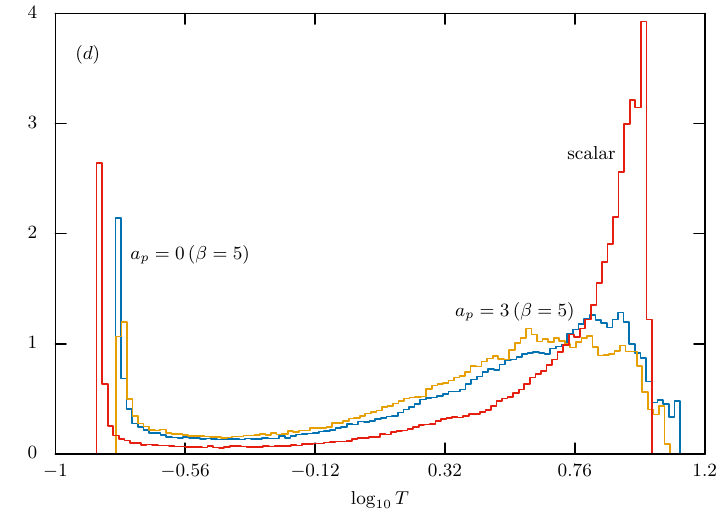}}\
    ~\vskip0pt
    \includegraphics[width=0.505\textwidth]{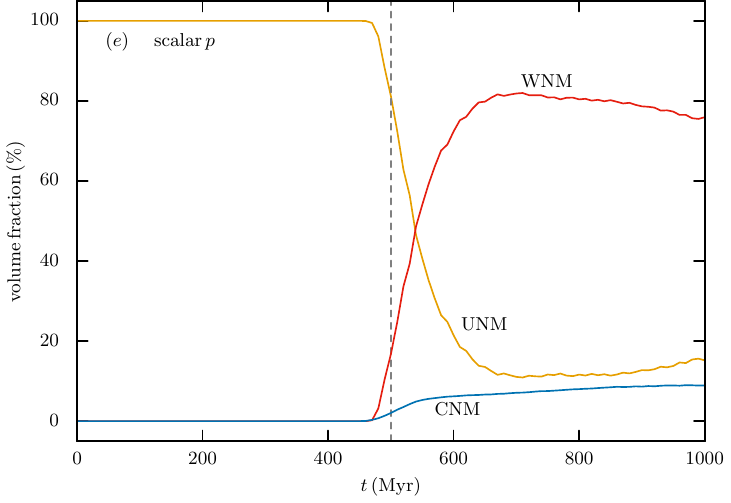}\hskip0pt\raisebox{-2pt}{\includegraphics[width=0.48\textwidth]{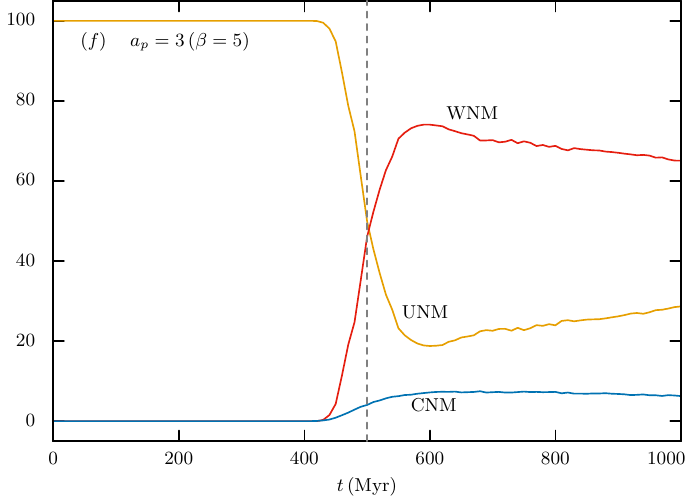}}
    \par
    %\end{centering}
    \caption{\label{fig:PDFs-for-}Mass-weighted density PDFs for $\beta=10^{4}$ [{panel (\emph{a})}] and $\beta=5$ [{panel (\emph{b})}]. {Panel (\emph{a})} also shows the {density} PDF for scalar pressure. {Panel (\emph{c})} shows the PDFs for pressure and {panel (\emph{d})} shows the PDFs for temperature. The bottom row shows the volume fraction $(\%)$ of the three phases across evolution timeline for scalar pressure [{panel (\emph{e})}] and for $\beta=5,a_{p}=3$ [{panel (\emph{f})}].}
\end{figure*}

In {Figs.\ref{fig:PDFs-for-}(\emph{a}) and (\emph{b})}, we have shown the probability density function (PDF) for the scalar pressure and anisotropic pressure with $\beta=5$ and $10^{4}$ and $a_{p}=0$ and $3$ at $850\,{\rm Myr}$. As we can see, without any ambient magnetic field, density distribution shows clear indication of two distinctly separated phases with a modest amount of unstable gas in the intermediate region. However, with a background magnetic field, the growth of condensation is restricted, leading to a wide dispersion in the lower density regions. While for `weak' magnetic field the condensations are roughly same for both $a_{p}=0$ and $3$, for the `strong' magnetic field, we see a marked difference between the final condensation distribution; specifically, when $p_{\parallel}$ is considerably higher than $p_{\perp}$, the thermal condensation is slightly suppressed, resulting in a greater proportion of unstable gas compared to the case when $p_{\parallel}\sim p_{\perp}$. Whereas the WNM mass fraction remain almost same $(16\% -17\%)$ for $a_{p}=0$ to $3$, CNM mass fraction reduces from $49\%$ to $42\%$ and in turn UNM mass fraction increases from $34\%$ to $41\%$. Detailed procedure of these analyses is mentioned in the following paragraph. {Fig.\ref{fig:PDFs-for-}(\emph{c})} shows the PDF of thermal pressure, from which we can see that pressure-redistribution is quite affected by the magnetic field making the turbulence even more isobaric.

\begin{figure*}
    %\begin{centering}
    %~\hskip4pt
    \includegraphics[width=0.495\textwidth]{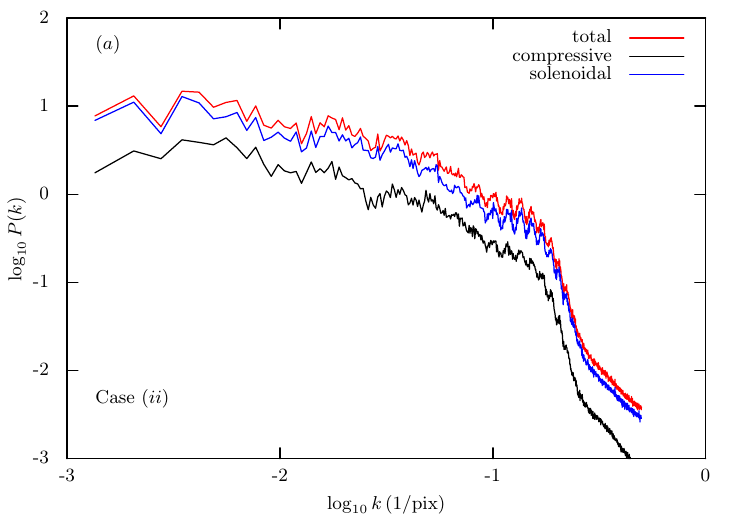}
    \includegraphics[width=0.495\textwidth]{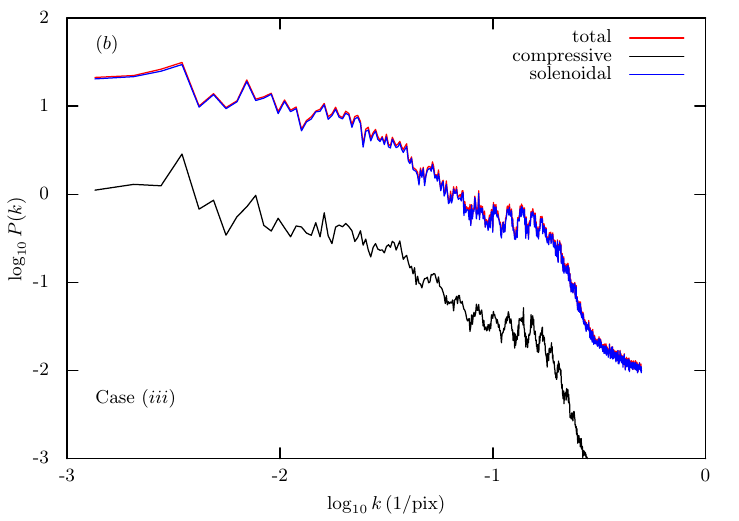}
    \par
    %\end{centering}
    \caption{\label{fig:sol1}Left: A velocity field with a dominant solenoidal component $(\sim72\%)$ and a minor compressive component $(\sim28\%)$, and Right: A predominantly solenoidal component $(\gtrsim95\%)$ and a negligible compressive component.}
\end{figure*}

We have also calculated the volume fraction occupied in the temperature PDFs, considering the unstable log--temperature range of $-0.7\,-\,0.25$, which corresponds to physical temperature $\sim100-890\,{\rm K}$. We find that minimum and maximum temperatures indicated in our simulation results lie between $\sim65-6400\,{\rm K}$, which are within the limits of observed temperatures for WNM, UNM, and CNM \citep{Heiles_2003a,Stahler_2004}. In the bottom row of Fig.\ref{fig:PDFs-for-}, we have shown the volume fraction $(\%)$ occupied over time for the three phases -- WNM, CNM, and UNM for scalar pressure [{Fig.\ref{fig:PDFs-for-}(\emph{e})}] and for $\beta=5\,\textrm{with}\,a_{p}=3$ [{Fig.\ref{fig:PDFs-for-}(\emph{f})}]. The dashed vertical lines in both panels indicate a transition of the TI from linear to nonlinear regime. We have calculated the volume and mass fractions occupied by these three phases at $t=850\,{\rm Myr}$ from the density PDFs. We consider the log--density range $\sim-0.25-0.7$ to be the unstable phase range which corresponds to density $\sim1-10\,{\rm cm^{-3}}$. Densities below this range are considered to be WNM and above are considered to be CNM. We find that while the volume fraction in WNM decreases from about $\sim80\%$ for scalar pressure to about $\sim68\%$ for $a_{p}=3$ with $\beta=5$, the same for CNM stays almost constant at about $\sim9-7\%$. Naturally, we have a considerable portion of the volume fraction in the UNM phase which increases from about $\sim11\%$ for scalar pressure to almost two times of that $\sim25\%$ for $a_{p}=3$ with $\beta=5$.

In contrast to the above, the mass fraction in the WNM phase stays almost constant at about $\sim16-17\%$ for both scalar and polybaric pressure, while that in the CNM phase decreases from about $\sim67\%$ for scalar pressure to about $\sim42\%$ for $a_{p}=3$ with $\beta=5$. The mass fraction in the UNM phase, however, increases by about $2.5$ times from $\sim17\%$ for scalar pressure to about $\sim41\%$ for $a_{p}=3$ with $\beta=5$. We note that all these findings are quite consistent with contemporary observational data, especially for the UNM phase \citep{Heiles_2003a,Kalberla_2018,Murray_2018}. So, the presence of a background magnetic field considerably affects the mass and volume distribution of the WNM and CNM phases, with the magnetic field making the condensations more diffused. Many previous studies have neglected analysis of the long-lasting UNM phase, considering it to be a transient phase formed between the WNM and CNM phases. However, our analysis shows that UNM occupies a very significant amount of fractions both volume and mass-wise and turbulence remains crucial in maintaining this substantial quantity of thermally unstable gas. Similar natures of filling factors were also reported by earlier simulations \citep{Kim_2023,Kritsuk_2017,Audit_2005}.

Regarding the distribution of size, we find that at $500\,{\rm Myr}$, the average size of cloud condensates without a magnetic field is $\sim(6\times2)\,{\rm pc^{2}}$ and for a background magnetic field, the average filament size is $\sim(14\times0.16)\,\text{pc}^{2}$ along the field direction. It is to be noted that these values are consistent with observations of ISM HI. For example, \citet{McClure-Griffiths_2006} observed a network of dozens of hair-like filaments of magnetically dominated cold hydrogen with lengths up to $\sim17\,{\rm pc}$ and widths $<0.1\,{\rm pc}$ while \citet{Clark_2014} have observed magnetically aligned HI fibres of length $\sim8.7\,\text{pc}$ and width $<0.12\,\text{pc}$ by studying diffuse galactic interstellar medium.

%%%%%%%%%%
%{
Our analysis  of the three ISM phases further confirms the non-transient nature of the UNM and the volume and mass fractions of the multiphase ISM, calculated from our simulation is consistent to the contemporary results \citep{Kim_2023,Kritsuk_2017,Audit_2005}, which indicates to the fact that with a polybaric pressure profile,  we arrive at a pretty much same results, so far as the structure of the multiphase ISM is concerned.  We  also note that the background magnetic field makes the size of the WNM (in terms of volume) less than the case with scalar pressure, at the cost of the UNM,  leaving the CNM size almost unchanged.
%}
%%%%%%%%%%

\begin{table}

\caption{\label{tab:comp}Comparison of final volume and mass fractions}
%~\vskip0pt

\begin{centering}
\begin{tabular}{|c|c|c|c|}
%\hline 
\multicolumn{4}{|c|}{Volume fractions (\%)\vrr}\tabularnewline
\hline 
Initial velocity field\vrr & UNM & CNM & WNM\tabularnewline
\hline 
\hline 
(\emph{i})\,\, $P_{{\rm com}}\sim54\%,P_{\rm sol}\sim46\%$\hfill~\vrr & $29$ & $6$ & $65$\tabularnewline
\hline 
(\emph{ii})\, $P_{{\rm sol}}\sim72\%,P_{{\rm com}}\sim28\%$ \hfill~\vrr & $35$ & $5$ & $60$\tabularnewline
\hline 
(\emph{iii}) $P_{{\rm sol}}\gtrsim95\%$\hfill~\vrr & $30$ & $6$ & $64$\tabularnewline
\hline 
\multicolumn{1}{c}{} & \multicolumn{1}{c}{} & \multicolumn{1}{c}{} & \multicolumn{1}{c}{}\tabularnewline
%\hline 
\multicolumn{4}{|c|}{Mass fractions (\%)\vrr}\tabularnewline
\hline 
\hline 
(\emph{i})\,\, $P_{{\rm com}}\sim54\%,P_{\rm sol}\sim46\%$\hfill~\vrr & $47$ & $36$ & $17$\tabularnewline
\hline 
(\emph{ii})\, $P_{{\rm sol}}\sim72\%,P_{{\rm com}}\sim28\%$\hfill~\vrr & $55$ & $26$ & $19$\tabularnewline
\hline 
(\emph{iii}) $P_{{\rm sol}}\gtrsim95\%$\hfill~\vrr & $48$ & $33$ & $19$\tabularnewline
\hline 
\end{tabular}
\par\end{centering}
\end{table}

\subsubsection{Effect of a solenoidal velocity field} \label{sec:6}

Finally, in this section, we present the comparative results of an initial velocity field with a dominant solenoidal component. We consider the other two cases, as mentioned in Section \ref{sec:5.1}. In Fig.\ref{fig:sol1}, we have shown the PSDs for these two cases -- Case\,(\emph{ii}): one with the ratio of $\sim72:28$ for the  solenoidal and compressive parts and Case\,(\emph{iii}): one with a dominant solenoidal part with only a negligible compressive component.

\begin{figure*}
    %\begin{centering}
    %~\hskip4pt
    \includegraphics[width=0.5\textwidth]{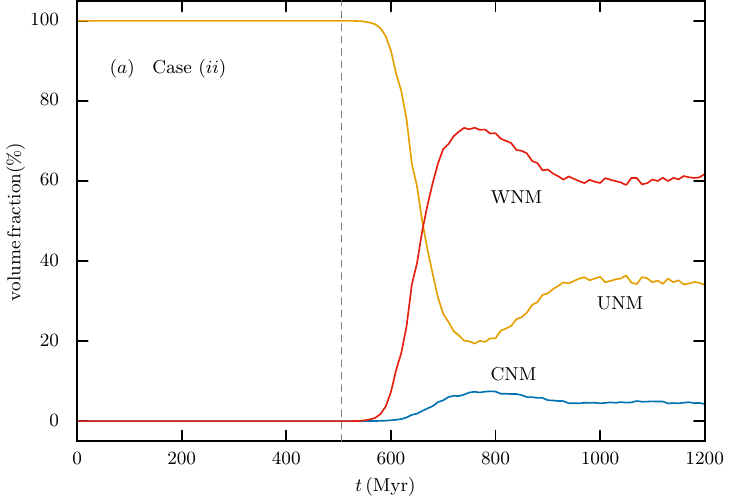}\hfill%
    \includegraphics[width=0.5\textwidth]{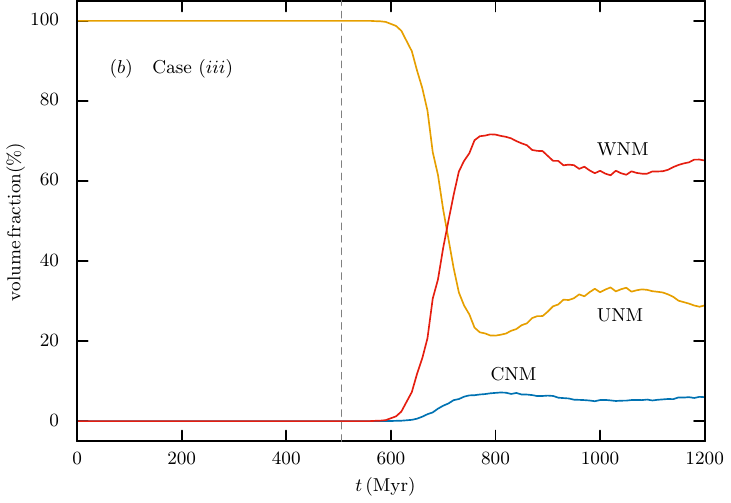}\\
    \includegraphics[width=0.5\textwidth]{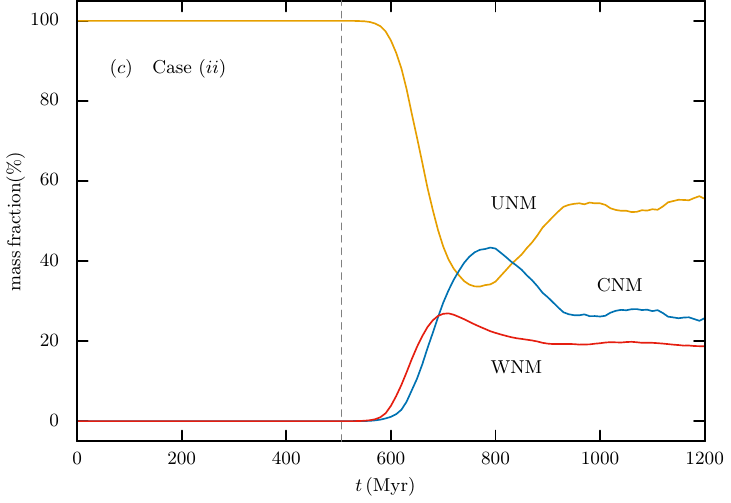}\hfill%
    \includegraphics[width=0.5\textwidth]{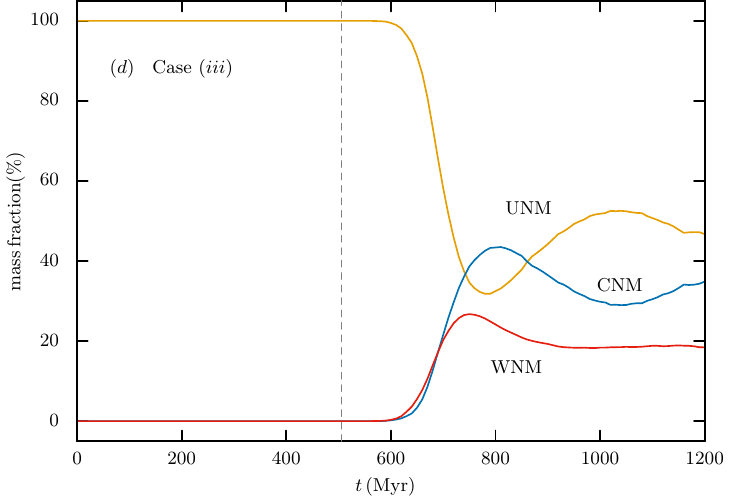}
    \par
    %\end{centering}
    \caption{\label{fig:sol-vol}Volume (top row) and mass (bottom row) fractions of multiphase ISM for an initial velocity field with  a dominant solenoidal component $(\sim72\%)$ and a minor component $(\sim28\%)$ { [panels (\emph{a}), (\emph{c})]}, and a predominantly solenoidal component $(\gtrsim95\%)$ and a negligible compressive component { [panels (\emph{b}), (\emph{d})]}. These two cases correspond to the corresponding velocity fields shown in Fig.\ref{fig:sol1}.}
\end{figure*}

The volume and mass fractions of the resultant multiphase ISM for an initial velocity field with dominant solenoidal component is shown in Fig.\ref{fig:sol-vol}. In the figure, the left and right panels, respectively correspond to the Cases (\emph{ii}) and (\emph{iii}) discussed in Section \ref{sec:5.1}. The top row {[Figs.\ref{fig:sol-vol}(\emph{a}, \emph{b})]} shows the volume fractions and the bottom row {[Figs.\ref{fig:sol-vol}(\emph{c}, \emph{d})]} shows the mass fractions. These volume fractions should be compared to the one with a velocity field with a relatively dominant compressive field shown in Fig.\ref{fig:PDFs-for-}. Apart from a relatively earlier starting time for growth of the nonlinear phase and subsequent saturation, statistically the volume and mass fractions of the WNM, UNM, and CNM for these two cases are similar to one with a relatively dominant compressive velocity field. The comparative values of the volume and mass fractions of these cases are shown in Table. \ref{tab:comp}, where the quantity $P_{\rm com, sol}$ denotes, respectively, the power in compressive and solenoidal parts of the initial velocity field. In all these cases, the volume and mass fractions are estimated during the quasi-steady state of the evolution when the nonlinear TI has saturated, which is about $\sim1000\,\rm Myr$ for Case (\emph{i}) and $\sim1150\,\rm Myr$ for Cases (\emph{ii}) and (\emph{iii}). From the table, we can infer that a more solenoidal initial velocity field (Case (\emph{ii})) favours a larger fraction of UNM at the cost of CNM and WNM, both mass and volume wise. Further increasing the solenoidal component in Case (\emph{iii}) reduces the mass and volume fractions, bringing them closer to their original values in Case (\emph{i}).

Regarding the timeline of development of the TI, we see that while for a relatively dominant compressive velocity field (with a marginally weak solenoidal part), the separation of the multiphase ISM starts around $450\,\rm Myr$, this begins at around $650-700\,\rm Myr$ for a velocity field with dominant solenoidal part. We  note that in all these cases, the average amplitude of the initial velocity field is same, so the late begin of the separation of ISM phases in the latter can be attributed to the fact that a compressive perturbation helps drive the TI making it reach its nonlinear phase in a comparatively short time. We \emph{must} however emphasize that the absolute time when the nonlinear phase begins which makes the beginning of the multiphase separation process is dependent on the initial perturbation strength -- the higher the initial perturbation strength, the quicker the TI reach its nonlinear saturation.
For the purpose of comparison, the mass fractions for the multiphase ISM for a relatively dominant compressive initial velocity field, a  case which we have already discussed in the previous sections, is shown in Fig.\ref{fig:sol0mass}. 

Analysis of the spatial PSD at the final stage of thermal instability reveals that the inertial-range turbulence scaling steepens from $E(k) \propto k^{-1.35}$ to $E(k) \propto k^{-1.43}$ as the initial velocity field changes from moderately solenoidal to predominantly solenoidal, respectively. The effects of this are evident in the multiphase segregation and mass and volume fractions discussed earlier. The steeper scaling observed in Case (\emph{iii}) suggests a more efficient energy cascade, resulting in enhanced phase segregation compared to Case (\emph{ii}), and finally leading to a higher CNM fraction.

\begin{table}
\caption{\label{tab:psd}Ratio of power $(P)$ in the initial and final velocity fields for different
cases.}

\centering{}%
\begin{tabular}{|c|c|c|}
\hline 
Case & Initial power ratio & Final power ratio\tabularnewline
 & $(P_{{\rm sol}}:P_{{\rm com}})$ & $(P_{{\rm sol}}:P_{{\rm com}})$\tabularnewline
\hline 
\hline 
(\emph{i}) & $46:54$ & $53:47$\tabularnewline
\hline 
(\emph{ii}) & $72:28$ & $52:48$\tabularnewline
\hline 
(\emph{iii}) & $95:5$ & $55:45$\tabularnewline
\hline 
\end{tabular}
\end{table}

We also note  the powers of velocity fields in the saturation stage of the instability, separate to an almost equal amounts in the solenoidal and compressive parts with a slightly more power in the solenoidal part, irrespective of the initial velocity field (see Table.\ref{tab:psd}).

\section{Summary and conclusion} \label{sec:7}

This study examines the nonlinear development of isobaric TI due to 
%%%%%%%%%%%%
%{
time-dependent
%} 
%%%%%%%%%%%%
radiative heat-loss in a collisionless warm anisotropic and inviscid plasma using a 2D FCT simulation, for a polybaric equation of state across three different scenarios: unmagnetized, weakly magnetized, and strongly magnetized. 
%The FCT simulation ensures that the simulation is not affected by numerical diffusion \citep{Boris_1973}. 
{The FCT simulation ensures that numerical diffusion remains minimised throughout the simulation while preventing unphysical oscillations \citep{Boris_1973}.} 
The simulation is started with a random velocity field with an adjustable compressive and solenoidal parts, which is supposed to be a result of supernov\ae\ forcing.  As time progresses, the TI starts to build up with the linear phase, before entering a quasilinear and finally a nonlinear saturation regime. As such, we establish the linear instability conditions for various modes along with their corresponding growth rates \citep{Field_1965}, which also serves as the benchmarking results for the nonlinear simulation. The effect of pressure anisotropy in the presence of an ambient magnetic field is handled with the polybaric equation of state \citep{Stasiewicz_2004}, helping us to maintain the polytropic nature of the thermal pressure. We also show that for the parameter regime considered in this work, kinematic viscosity can be safely neglected. Apparently, our simulation with the relevant parameters, is able to resolve all the necessary time and length scales.

The important findings of our study can be summarized as:
\begin{itemize}
\item In the absence of background magnetic field, a relatively dominant compressive forcing drives the turbulence to Burger like inertial range scaling but with relatively more power in solenoidal component of the PSD. The nature of the turbulence remains subsonic.

\item For the `strong' magnetic field with relatively dominant compressive forcing, the turbulence scales like the classical Kolmogorov turbulence, suggesting a transition toward an incompressible, subsonic turbulent regime. We also note similar kind of low frequency, isobaric, incompressible turbulence in the VLISM, as reported recently by \citet{Zhao_2020}. We have also observed that, for moderately dominant solenoidal forcing, phase segregation relatively gets suppressed a little due to inefficient turbulence cascading as confirmed by a flatter inertial range scaling in PSD. In contrast to this, under predominantly solenoidal forcing, the relatively steeper inertial-range scaling, approaching the Kolmogorov $-5/3$ slope, suggests a more efficient energy cascade leading to a higher fraction of CNM in the final stage.

\item At the nonlinear saturation regime of the TI, our analysis also shows the ISM  to be consisting of three-phase matter, namely WNM, CNM, and UNM. The volume and mass fractions of the three ISM phases are consistent with the contemporary studies which include external forcing and star formation activities \citep{Kim_2023,Kritsuk_2017}.

\begin{figure*}
    \begin{centering}
    %~\hskip4pt
    \includegraphics[width=0.5\textwidth]{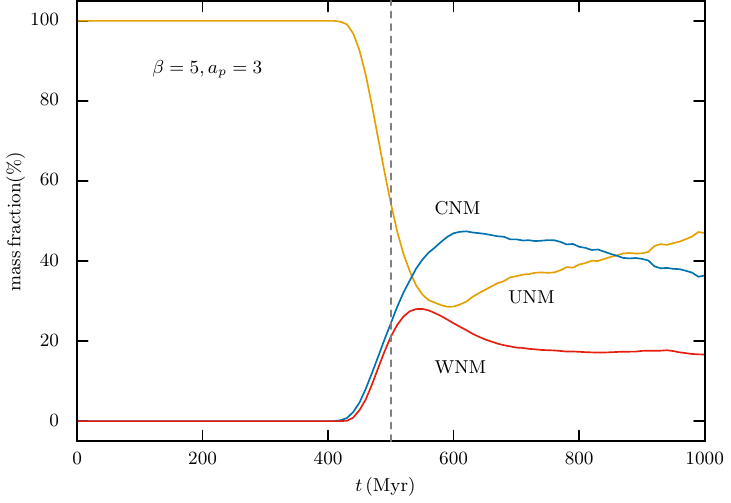}
    \end{centering}
    \caption{\label{fig:sol0mass}The mass fractions of multiphase ISM for a relatively dominant compressive initial velocity field. This for the case that we have already discussed in Section \ref{sec:5.2}. This case is for $\beta=5$ and polybaric parameter $a_p=3$. The corresponding volume fraction is shown {in Fig.\ref{fig:PDFs-for-}(\emph{f})}.}
\end{figure*}

\item The presence of a homogenous background magnetic field severely affects the formation of condensations, mainly restricting the fluid flow in the perpendicular direction and thereby elongating the condensates to form filaments along the direction of the magnetic field with an average length of about $\sim14\,\text{pc}$ and width $\sim0.16\,\text{pc}$, which closely agrees with existing observational studies \citep{McClure-Griffiths_2006,Clark_2014}.

\item As far as the effect of polybaric pressure anisotropy is concerned, we have observed that the growth of TI is suppressed when the parallel component of the pressure becomes larger than the perpendicular component, leading to a comparatively larger fraction of the UNM. This is indicative of the fact that the anisotropy parameter $a_{p}$ plays a key role in the evolution of the ISM. Obviously, the effect of pressure anisotropy increases with the increase in magnetic field strength. For both `weak' and `strong' fields, the higher is the value of $a_{p}$, the lower is the mass and volume fraction of the condensation. However, the selected magnetic field strengths do not exhibit any significant differences in the overall evolution, apart from a small enhancement of anisotropic effects. We would also like to note that a stronger field is expected to cause defragmentation of the field aligned strands, which may be expected at a much lower $\beta$ for our case \citep{Choi_2012}.   

{\item The formation of condensations is strongly influenced by both thermal conduction and pressure anisotropy. Without thermal conduction, filamentary structures tend to form across the direction of magnetic field \citep{wareing_2016}, whereas the presence of conduction aligns the condensations along the magnetic field direction. Additionally, stronger magnetic fields enhance the effects of pressure anisotropy. Higher anisotropy leads to filaments getting elongated along the direction transverse to the magnetic field, while lower anisotropy results in more scattered, nearly isotropic structures.}
\end{itemize}
%As a concluding remark, we would like to emphasize that, to our knowledge, this study is the first-ever comprehensive numerical study of TI in the ISM using a flux-corrected transport (FCT) method and also the first study to use polybaric pressure model in investigating TI in the ISM.
{As a concluding remark, we would like to emphasize that, to our knowledge, this study is the first-ever comprehensive numerical study to use polybaric pressure model in investigating TI in the ISM, thereby offering new insights regarding the effects of pressure anisotropy on multiphase structure formation and we believe it establishes a foundation for future investigations, which should address the case of tangled magnetic fields evolution for a more realistic system.}

\section*{Acknowledgements}

One of the authors, HS would like to thank CSIR-HRDG, New Delhi, India for the Senior Research Fellowship grant no: 09/059(0074)/2021-EMR-I. {The authors to like to thank the referee for constructive suggestions.}

%%%%%%%%%%%%%%%%%%%%%%%%%%%%%%%%%%%%%%%%%%%%%%%%%%
\section*{Data Availability}
The data that support the findings of this study are available within the article. 



\bibliographystyle{mnras}
\bibliography{ReferencesTI} % if your bibtex file is called example.bib

%\begin{thebibliography}{99}
%\end{thebibliography}

%%%%%%%%%%%%%%%%%%%%%%%%%%%%%%%%%%%%%%%%%%%%%%%%%%

%%%%%%%%%%%%%%%%% APPENDICES %%%%%%%%%%%%%%%%%%%%%

%\appendix


%If you want to present additional material which would interrupt the flow of the main paper, it can be placed in an Appendix which appears after the list of references.

%%%%%%%%%%%%%%%%%%%%%%%%%%%%%%%%%%%%%%%%%%%%%%%%%%

% Don't change these lines
\bsp	% typesetting comment
\label{lastpage}
\end{document}